\documentclass[]{aa}
\usepackage{epsf,palatino,url,graphicx,wrapfig,array,setspace,amsmath,amssymb,fancyhdr,multirow,lscape,appendix,rotating,txfonts}
\usepackage{graphics,epsfig,txfonts}
\usepackage{natbib}
\usepackage{subcaption}
\bibpunct{(}{)}{;}{a}{}{,}

\usepackage{graphicx}
\usepackage[flushleft]{threeparttable}
\usepackage{booktabs} 

\usepackage{longtable}
\usepackage{amssymb}
\usepackage{xcolor}
\usepackage{color}
\usepackage{lipsum}
\usepackage{textcomp}
\usepackage[colorlinks=true,linkcolor=blue,citecolor=blue,
bookmarks=true,bookmarksopen=true,bookmarksnumbered=true,draft=false]{hyperref}
\usepackage{enumerate}

\usepackage[normalem]{ulem}
\usepackage{soul}
\usepackage[switch]{lineno}
\usepackage{blindtext}

\usepackage{epsf}
\usepackage{amssymb}
\usepackage{epsfig}
\usepackage{pdflscape}
\usepackage{amsmath}
\usepackage{graphicx}
\usepackage{pdflscape}

\newcommand{\ergs}[1]{$\times 10^{#1}$ erg s$^{-1}$}
\newcommand{\oergs}[1]{$10^{#1}$ erg s$^{-1}$}

\newcommand{\ct}{cts s$^{-1}$}

\newcommand{\ltsima}{$\buildrel < \over \sim$}
\newcommand{\lsim}{\lower.5ex\hbox{\ltsima}}
\newcommand{\gtsima}{$\buildrel > \over \sim$}
\newcommand{\gsim}{\lower.5ex\hbox{\gtsima}}

\newcommand{\sxp}{SMC\,X-3\xspace}
\newcommand{\swift}{{\it Swift}\xspace}

\newcommand{\xmm}{{\it XMM-Newton}\xspace}

\newcommand{\cxo}{{\it Chandra}\xspace}

\begin{document}

\title{Accreting, highly magnetized neutron stars at the Eddington limit: A study of the 2016 outburst of SMC X-3}

\author{Filippos Koliopanos\inst{1,2}\thanks{\email{fkoliopanos@irap.omp.eu}}
 \and  Georgios Vasilopoulos\inst{3}}

\titlerunning{SMC X-3 during outburst}
\authorrunning{Koliopanos et al.}

\institute{CNRS, IRAP, 9 Av. colonel Roche, BP 44346, F-31028 Toulouse cedex 4, France
          \and Universit{\'e} de Toulouse; UPS-OMP; IRAP, Toulouse, France 
          \and Max-Planck-Institut für Extraterrestrische Physik, Giessenbachstraße, 85748 Garching, Germany}

\date{Received ... / Accepted ...}

  \abstract
   {}
   {We study the temporal and spectral characteristics of \sxp during its recent (2016) outburst to probe accretion onto highly magnetized neutron stars (NSs) at the Eddington limit.}
   {We obtained \xmm observations of \sxp and combined them with long-term observations by \swift. We performed a detailed analysis of the temporal and spectral behavior of the source, as well
as its short- and long-term evolution. We have also constructed a simple toy-model (based on robust theoretical predictions)  in order to gain insight into the complex emission pattern of \sxp.}
   {We confirm the pulse period of the system that has been derived by previous works and note that the pulse  has a complex three-peak shape. We find that the pulsed emission is dominated by hard photons, while at energies below ${\sim}1$\,keV, the emission does not pulsate. We furthermore find that the shape of the pulse profile and the short- and long-term evolution of the source light-curve can be explained by invoking a combination of a ``fan'' and a ``polar'' beam. The results of our  temporal study are supported by our spectroscopic analysis, which reveals a two-component emission, comprised of a hard power law and a soft thermal component. We find that the latter produces the bulk of the non-pulsating emission and is most likely the result of reprocessing the primary hard emission by optically thick material that partly obscures the central source.  We also detect strong emission lines from highly ionized metals. The strength of the emission lines strongly depends on the phase.}
   {Our findings are in agreement with previous works. The energy and temporal evolution as well as the shape of the pulse profile and the long-term spectra evolution of the source are consistent with the expected emission pattern of the accretion column in the super-critical regime, while the large reprocessing region is consistent with the analysis of previously studied X-ray pulsars observed at high accretion rates. This reprocessing region is consistent with recently proposed theoretical and observational works that suggested that highly magnetized NSs occupy a considerable fraction of ultraluminous X-ray sources. }
   {}

\maketitle

\section{Introduction}
\label{sec-intro}

X-ray pulsars (XRPs) are comprised of a  highly magnetized ($B{>}10^9$\,G) neutron star (NS) and a companion star that ranges from low-mass white dwarfs to massive B-type stars \citep[e.g.,][and references therein]{2012MmSAI..83..230C,2015A&ARv..23....2W,2016arXiv160806530W}. XRPs are some of the most luminous  (of-nuclear) Galactic X-ray point-sources \citep[e.g.,][]{2017Sci...355..817I}. The X-ray emission is the result of accretion of material from the  star onto the NS, which in the case of XRPs is  strongly affected by the NS magnetic field: The accretion disk formed by the in-falling matter from the companion star is truncated at approximately the NS magnetosphere. From this point, the accreted material flows toward the NS magnetic poles, following the magnetic field lines, forming an  accretion column, inside which material is heated to high energies \citep[e.g.,][]{1975A&A....42..311B,1976MNRAS.175..395B,1985ApJ...299..138M}. The opacity inside the accretion column is dominated by scattering between photons and electrons. In the presence of a high magnetic field, the scattering cross-section is highly anisotropic \citep{1971PhRvD...3.2303C,1974ApJ...190..141L}, the hard X-ray photons from the accretion column are collimated in a narrow beam (so-called pencil beam), which is directed parallel to the magnetic field axis \citep{1975A&A....42..311B}. The (possible) inclination  between the pulsar beam and the rotational axis of the pulsar, combined with the NS spin, produce the characteristic pulsations of the X-ray light curve.

During episodes of prolonged accretion, XRPs are known to reach and often exceed the Eddington limit for spherical accretion onto a NS (${\sim}1.8\times10^{38}$\,erg/s for a $1.4\,{\rm M_{\odot}}$ NS).
This is further complicated by the fact that material is accreted onto a very small area on the surface of the NS, and as a result, the Eddington limit is significantly lower. Therefore, even  X-ray pulsars with luminosities of ${{\text{about }}}$a\,few\,$10^{37}$\,erg/s persistently break the (local) Eddington limit.  \cite{1976MNRAS.175..395B} demonstrated that while the accreting material is indeed impeded by the emerging radiation and the accretion column becomes opaque along the magnetic field axis, the X-ray photons can escape from the (optically thin) sides of the accretion funnel in a ``fan-beam'' pattern \citep[see, e.g., Fig.1 in][]{2007A&A...472..353S} that is directed perpendicular to the magnetic field. More recent considerations also predict that a fraction of the fan-beam emission can be reflected off the surface of the NS, producing a secondary polar beam that is directed parallel to the magnetic field axis \citep{2013ApJ...764...49T,2013ApJ...777..115P}. 

The recent discoveries of pulsating ultraluminous X-ray sources (PULXs) have established that XRPs can persistently emit at luminosities that are hundreds of times higher than the Eddington luminosity \citep[e.g.,][]{2014Natur.514..202B,2016ApJ...831L..14F,2017Sci...355..817I}. Refinements of the mechanism reported by \cite{1976MNRAS.175..395B} demonstrated that accreting highly magnetized NSs can facilitate luminosities exceeding $10^{40}$\,erg\,${\rm s^{-1}}$ \citep[][]{2015MNRAS.454.2539M}, while other recent studies propose that many (if not most) ULXs are accreting (highly magnetized) NSs, rather than black holes \citep{2016MNRAS.458L..10K,2017MNRAS.468L..59K,2017ApJ...836..113P,2017MNRAS.467.1202M, kolio2017}.

Understanding the intricacies of the physical mechanisms that underly accretion onto high-B NSs requires the detailed spectral and timing analysis of numerous XRPs during high accretion episodes. The shape of the pulse profile in different energy bands and at different luminosities provides valuable insights into the shape of the emission pattern of the accretion column and may also shed light on the geometry and size of the accretion column itself \citep[e.g.,][]{1981ApJ...251..288N, 1983ApJ...270..711W,1992hrfm.book.....M, 1996BASI...24..729P, 1997A&A...319..507P, 2004A&A...421..235R, 2013A&A...558A..74V,2014A&A...567A.129V,2016MNRAS.456.3535K}. Furthermore, the shape of the spectral continuum, its variability, and the fractional variability of the individual components of which it is comprised, along with the presence (or absence), shape and variability of emission-like features are directly related to the radiative processes that are at play in the vicinity of the accretion column, and may also reveal the presence of material trapped inside or at the boundary of the magnetosphere (\citealt{1980A&A....87..330B}; \citealt{1980A&A....86..121S}; \citealt{2002ApJ...564L..21S}; \citealt{2006A&A...455..283L}; \citealt{2012MNRAS.425..595R}; \citealt{2016MNRAS.461.1875V}; Vasilopoulos et al. 2017 in press).

The emission of the accretion column of XRPs has a spectrum that is empirically described by a very hard power law (spectral index ${\lesssim}1.8$) with a low-energy (${\lesssim}10$\,keV) cutoff \citep[e.g.,][]{2012MmSAI..83..230C}. It has been demonstrated that the spectrum can be reproduced  assuming bulk and thermal Comptonization of bremsstrahlung, blackbody, and cyclotron seed photons \citep{1981ApJ...251..288N,1985ApJ...299..138M,1991ApJ...367..575B,2004ApJ...614..881H,2007ApJ...654..435B}. Nevertheless, a comprehensive, self-consistent modeling of the emission of the accretion column has not been achieved so far. The spectra of XRPs often exhibit a distinctive spectral excess below ${\sim}1$\,keV, which is successfully modeled using a blackbody component at a temperature of ${{\sim}}$0.1-0.2\,keV \citep[e.g.,][and references therein]{2004ApJ...614..881H}. While some authors have argued that  this ``soft excess'' is the result of Comptonization of seed photons from the truncated disk by low-energy ($kT_{e}{\lesssim}1$\,keV) electrons \citep[e.g.,][]{2001ApJ...553..375L}, others have noted that the temperature of the emitting region is hotter and its size considerably larger than what is expected for the inner edge of a truncated \cite{1973A&A....24..337S} accretion disk, and they attribute the feature to reprocessing  of hard X-rays  by optically thick material, trapped at the boundary of the magnetosphere. More recently, \cite{2017MNRAS.467.1202M} have extended these arguments to the super-Eddington regime, arguing that at high mass accretion rates (considerably above the Eddington limit), this optically thick material engulfs the entire magnetosphere and reprocesses most (or all) of the primary hard emission. The resulting accretion ``curtain'' is substantially hotter (${{\gtrsim}}1\,$keV) than that of the soft excess in the sub-Eddington sources.

Be/X-ray binaries \citep[BeXRBs; for a recent review, see][]{2011Ap&SS.332....1R} are a subclass of high-mass X-ray binaries (HMXB) that contains the majority of the known accreting X-ray pulsars with a typical magnetic field strength equal to or above $3\times10^{11}$~G \citep{2015MNRAS.454.3760I,2016ApJ...829...30C}.
In BeXRBs, material is provided by a non-supergiant Be-type donor star. This is a young stellar object that rotates with a near-critical rotation velocity, resulting in strong mass loss via an equatorial decretion disk \citep{2011A&A...527A..84K}.
As BeXRBs are composed of young objects, their number within a galaxy correlates with the recent star formation \citep[e.g.,][]{2010ApJ...716L.140A,2016MNRAS.459..528A}. 
By monitoring BeXRB outbursts, it is now generally accepted that normal or so-called Type I outbursts ($L_x\sim$\oergs{36})  can occur as the NS passes close to the decretion disk, thus they appear to be correlated with the binary orbital period. Giant or Type II outbursts ($L_{x}\ge$\oergs{38}) that can last for
several orbits are associated with wrapped Be-disks \citep{2013PASJ...65...41O}. 
From an observational point of view, the former can be quite regular, appearing in each orbit \citep[e.g., LXP\,38.55 and EXO\,2030+375;][]{2016MNRAS.461.1875V,2008ApJ...678.1263W}, while for other systems, they are rarer \citep{2015int..workE..78K}. On the other hand, major outbursts producing \oergs{38} are rare events with only a hand-full detected every year. 
Moreover, it is only during these events that the pulsating NS can reach super-Eddington luminosities, thus providing a link between accreting NS and the emerging population of NS-ULXs \citep{2014Natur.514..202B,2016ApJ...831L..14F,2017Sci...355..817I,2017MNRAS.466L..48I}.

The spectroscopic study of BeXRB outbursts in our Galaxy is hampered by the strong Galactic absorption and the often large uncertainties in their distances.  
The Magellanic Clouds (MCs) offer a unique laboratory for studying BeXRB outbursts.  
The moderate and well-measured distances of $\sim$50~kpc for the Large Magellanic Cloud (LMC) \citep[][]{2013Natur.495...76P} and $\sim$62~kpc for the Small Magellanic Cloud (SMC) \citep[][]{2014ApJ...780...59G}
as well as their low Galactic foreground absorption ($\sim$6$\times10^{20}$cm$^{-2}$) make them ideal targets for this task.

The 2016 major outburst of \sxp offers a rare opportunity to study accretion physics onto a highly magnetized NS during an episode of high mass accretion.
\sxp was one of the earliest X-ray systems to be discovered in the SMC in 1977 by the SAS~3 satellite at a luminosity level of about \oergs{38} \citep{1978ApJ...221L..37C}.
The pulsating nature of the system was revealed more than two decades later, when 7.78 s pulsations were measured \citep{2003HEAD....7.1730C} using data from the Proportional Counter Array on board the Rossi X-ray Timing Explorer \citep[PCA, RXTE;][]{2006ApJS..163..401J}, and a proper association was possible through an analysis of archival \cxo observations \citep{2004ATel..225....1E}.
By investigating the long-term optical light-curve of the optical counterpart of \sxp, \citet{2004AJ....128..709C} reported a 44.86~d modulation that they interpreted as the orbital period of the system. 
\citet{2008MNRAS.388.1198M} have reported a spectral type of B1-1.5 IV-V for the donor star of the BeXRB.

The 2016 major outburst of \sxp was first reported by MAXI \citep{2016ATel.9348....1N} and is estimated to have started around June 2016 \citep{2016ATel.9362....1K}, while the system still remains active at the time of writing (June 2017). Lasting for more than seven binary orbital periods, the 2016 outburst can be classified as one of the longest ever observed for any BeXRB system.  
Since the outburst was reported, the system has been extensively monitored in the X-rays by \swift with short visits, while deeper observations have been performed by NuSTAR, \cxo, and \xmm. 
\citet{2017arXiv170102336T} have reported a 44.918~d period as the true orbital period of the system by modeling the pulsar period evolution (see also \citealt{2017ApJ...843...69W}) during the 2016 outburst with data obtained by \swift/XRT and by analyzing the optical light-curve of the system, obtained from the Optical Gravitational Lensing Experiment \citep[OGLE,][]{2015AcA....65....1U}.
The maximum luminosity obtained by \swift/XRT imposes an above-average maximum magnetic field strength at the NS surface ($>10^{13}$~G), as does the lack of any cyclotron resonance feature in the high-energy X-ray spectrum of the system \citep{2016ATel.9404....1P,2017arXiv170200966T}. On the other hand, the lack of a transition to the propeller regime during the evolution of the 2016 outburst suggests a much weaker magnetic field at the magnetospheric radius. According to \citet{2017arXiv170200966T},  this apparent contradiction  could be resolved if there were a significant non-bipolar component of the magnetic field close to the NS surface.

Following the evolution of the outburst, we requested a ``non anticipated'' \xmm ToO observation (PI: F.~Koliopanos) in order to perform  a detailed study of the soft X-ray spectral characteristics of the system. In Sect.~\ref{sec-observations} we describe the spectral and temporal analysis of the \xmm data. In Sect.~\ref{sec:irr} we introduce a toy model constructed to phenomenologically explain our findings, and in Sects.~\ref{discussion} and~\ref{conclusion} we discuss our findings and interpret them in the context of highly accreting X-ray pulsars in (or at the threshold of) the ultraluminous regime.

\section{Observations and data analysis}
\label{sec-observations}

\subsection{\xmm data extraction}

\xmm observed SMC X-3 on October 14, 2016 for a duration of $\sim37$\,ks. All onboard instruments were operational, with MOS2 and pn  operating in Timing Mode and MOS1 in Large Window Mode. All detectors had the Medium optical blocking filter on. In Timing mode, data are registered in one dimension, along the column axis. This results in considerably shorter CCD readout time, which increases the
spectral resolution, but also protects observations of bright sources from pile-up\footnote{For more information on pile-up, see http://xmm2.esac.esa.int/docs/documents/CAL-TN-0050-1-0.ps.gz}.

Spectra from all detectors were extracted using the latest XMM-Newton Data Analysis software SAS, version 15.0.0., and using the calibration files released\footnote{XMM-Newton CCF Release Note: XMM-CCF-REL-334}  on May 12, 2016. The observations were inspected for high background flaring activity by extracting high-energy light curves (E$>$10\,keV for MOS and 10$<$E$<$12\,keV for pn) with a 100\,s bin size. The examination revealed $\sim$2.5\,ks of the pn data that where contaminated by high-energy flares. The contaminated intervals where subsequently removed. The spectra were extracted using
the SAS task \texttt{evselect}, with the standard filtering flags (\texttt{\#XMMEA\_EP \&\& PATTERN<=4} for pn and \texttt{\#XMMEA\_EM \&\& PATTERN<=12} for MOS). The SAS tasks \texttt{rmfgen} and \texttt{arfgen} were used to create the redistribution matrix and ancillary file. The MOS1 data suffered from pile-up and were
therefore rejected. For simplicity and self-consistency, we opted to only use the EPIC-pn data in our spectral analysis. The pn spectra are optimally calibrated for spectral fitting and had $\sim$1.3$\times10^6$ registered counts, allowing us to accurately constrain emission features and minute spectral variations between spectra extracted during different pulse phases.  While an official estimate is not provided by the \xmm SOC, the pn data in timing mode are expected to suffer from systematic uncertainties in the  1-2\% range (e.g., see Appendix A in \citealt{2014A&A...571A..76D}). Accordingly, we quadratically added 1\% systematic errors to each energy bin of the pn data.
The spectra from the Reflection Grating Spectrometer (RGS, \citealt{2001A&A...365L...7D}) were extracted using SAS task {\texttt{rgsproc}. The resulting RGS1 and RG2 spectra were combined using {\texttt rgscombine}.

\subsection{\swift data extraction}

To study the long-term behavior of the source hardness, pulse fraction, and evolution throughout the 2016 outburst, we also analyzed the available \swift/XRT data up to December 31, 2016. 
The data were analyzed following the instructions described in the \swift data analysis guide\footnote{\url{http://www.swift.ac.uk/analysis/xrt/}} \citep{2007A&A...469..379E}.
We used {\tt xrtpipeline} to generate the \swift/XRT products, while events were extracted by using the command line interface {\tt xselect} available through HEASoft FTOOLS \citep{1995ASPC...77..367B}\footnote{\url{http://heasarc.gsfc.nasa.gov/ftools/}}. 
Source events were extracted from a 45\arcsec region, while background was extracted from an annulus between 90\arcsec and 120\arcsec.

\subsection{X-ray timing analysis}
\label{time}

We searched for a periodic signal in the barycentric corrected EPIC-pn event arrival time-series by using an epoch-folding technique \citep[EF;][]{1990MNRAS.244...93D,1996A&AS..117..197L}.
The EF method uses a series of trial periods within an appropriate range to phase-fold the detected event arrival times and performs a $\chi^2$ minimization test of a constant signal hypothesis. 
Folding a periodic light curve with an arbitrary period smears out the signal, and the folded profile is expected to be nearly flat.
Thus a high value of the $\chi^2$ (i.e., a bad fit) indirectly supports the presence of a periodic signal.
A disadvantage of this method is that it lacks a proper determination of the period uncertainty. 
In many cases, the FWHM of the $\chi^2$ distribution is used as an estimate for the uncertainty, but this value only improves with the length of the time series and not with the number of events \citep{1996ApJ...473.1059G} and does not have the meaning of the statistical uncertainty. 
In order to have a correct estimate for both the period and its error, we applied
the Gregory-Loredo method of Bayesian periodic signal detection \citep{1996ApJ...473.1059G}, and constrained the search around the true period derived from the epoch-folding method.
To further test the accuracy in the derived pulsed period, we performed 100 repetitions of the above method while bootstrapping the event arrival times and by selecting different energy bands with 1.0 keV width (e.g., 2.0-3.0 keV and 3.0-4.0 keV). The derived period deviation between the above samples was never above $10^{-5}$~sec. 
From the above treatment, we derived a  best-fit pulse period of 7.77200(6)~s. The pulse profile corresponding to the best-fit period is plotted in Fig. \ref{fig:pp}. 
The accurate calculation of the pulse profile enabled us to compute the pulsed fraction (PF), which is defined as the ratio between the difference and the sum of the maximum and minimum count rates over the pulse profile, i.e.,  
\begin{equation}
PF=(F_{max}-F_{min})/(F_{max}+F_{min}). 
\label{pfeq}
\end{equation}
For the \xmm EPIC-pn pulse profile (Fig. \ref{fig:pp}) we calculated a pulsed fraction of $PF=0.328\pm0.005$.
 We also estimated the PF for five energy bands, henceforth noted with the letter $i$, with $i=1,2,3,4,5$ and
$1\rightarrow(0.2-0.5)$ keV, $2\rightarrow(0.5-1.0)$ keV, $3\rightarrow(1.0-2.0)$ keV, $4\rightarrow(2.0-4.5)$ keV, $5\rightarrow(4.5-10.0)$ keV \citep{2009A&A...493..339W,2013A&A...558A...3S}. The resulting values for the energy-resolved PF are $PF_1=0.208$, $PF_2=0.249$, $PF_3=0.341$, $PF_4=0.402$ and $PF_5=0.410$.
\begin{figure}
   \resizebox{\hsize}{!}{\includegraphics[angle=0,clip=,bb=15 31 522 500]{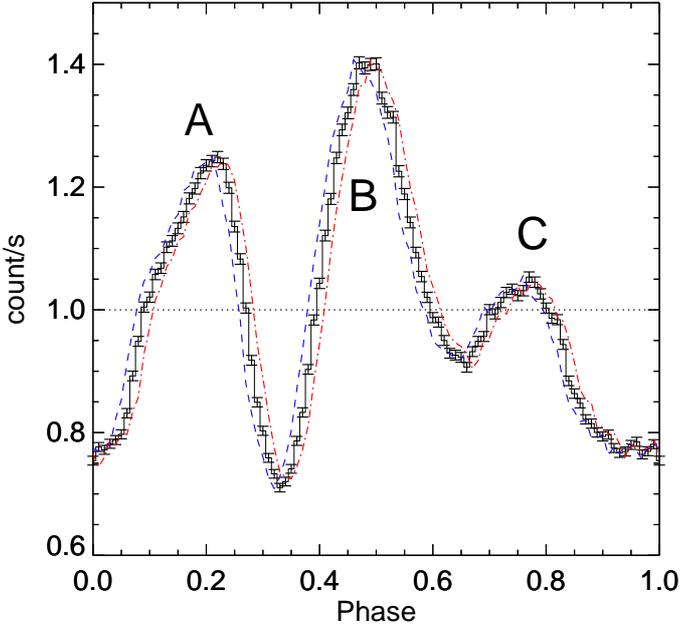}}
  \caption{Phase-folded pulse profile of \sxp derived from the best-fit period of 7.7720059. The profile is normalized at an average count rate of 44.49 \ct. Red and blue lines correspond to the derived pulse profile for a period shifted by $\pm5\times10^{-5}$~s, which is $\sim$1000 times larger than the derived period uncertainty. This was chosen to demonstrate the accuracy of the derived period. Three peaks  are clearly distinguished.}
  \label{fig:pp}
\end{figure}

\begin{figure}
   \resizebox{\hsize}{!}{\includegraphics[angle=0,clip=,bb=31 49 483 500]{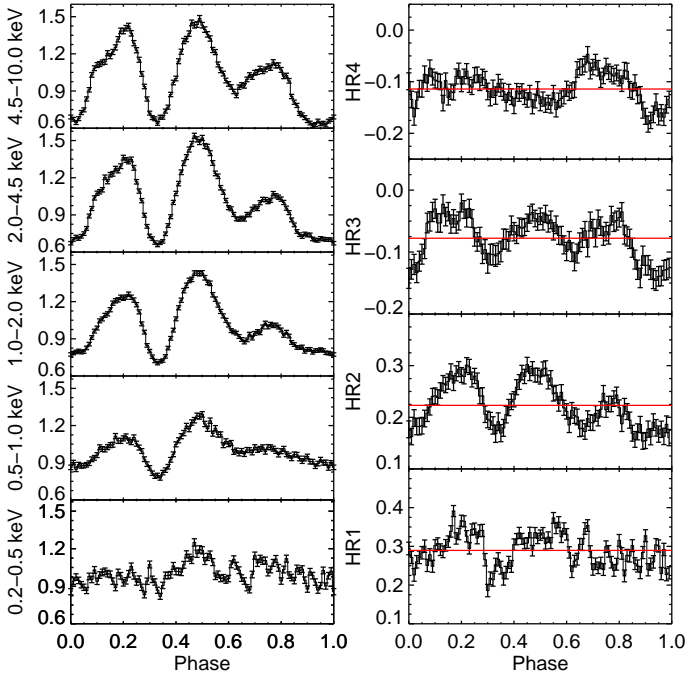}}
  \caption{
     Left: Pulse profiles of \sxp obtained from the EPIC-pn data in different energy bands. 
   The profiles are background subtracted and normalized to their 
   average count rate (from bottom to top: 4.33, 7.90, 12.6, 10.9, and 8.68 \ct).
   Right: Hardness ratios derived 
   from the pulse profiles in two neighboring standard energy bands as a function of pulse phase.
   The red horizontal lines denote the pulse-phase-averaged HR.
   }   
  \label{fig:HR}
\end{figure}

In Fig.~\ref{fig:HR} we present the phase-folded light curves for the five energy bands introduced above, together with the four phase-resolved hardness ratios.
The hardness ratios $HR_i$ ($i=1,2,3,4$) are defined as
\begin{equation}
HR_i = \frac{R_{i+1} - R_{i}}{R_{i+1} + R_{i}}
,\end{equation}
with  R$_{\rm i}$ denoting the background-subtracted count rate in the i$^{\rm th}$ energy band.
The plots revealed that the prominence of the three major peaks are strongly dependent on the energy range: the pulsed emission appears to be dominated by high-energy photons (i.e., $>2$\,keV), while in the softer bands, the pulse shape is less pronounced.
To better visualize the spectral dependence of the pulse phase, we produced a 2D histogram for the number of events as function of energy and phase (top panel of Fig.~\ref{fig:phase_heat}). 
The histogram of the events covers a wide dynamic range because
the efficiency of the camera and telescope varies strongly with energy.
To produce an image where patterns and features (such as the peak of the pulse) are more easily recognized, we normalized the histogram by the average effective area of the detector for the given energy bin and then normalized it a second time by the phase-averaged count rate in each energy bin (middle panel, Fig.~\ref{fig:phase_heat}).
Last, in the lower bin, each phase-energy bin was divided by the average soft X-ray flux (0.2-0.6 keV) of the same phase bin (bottom panel, Fig.~\ref{fig:phase_heat}). This energy range was chosen because it displays  minimum phase variability (Fig.~\ref{fig:HR}).

\begin{figure}
   \resizebox{\hsize}{!}{\includegraphics[angle=0,clip=]{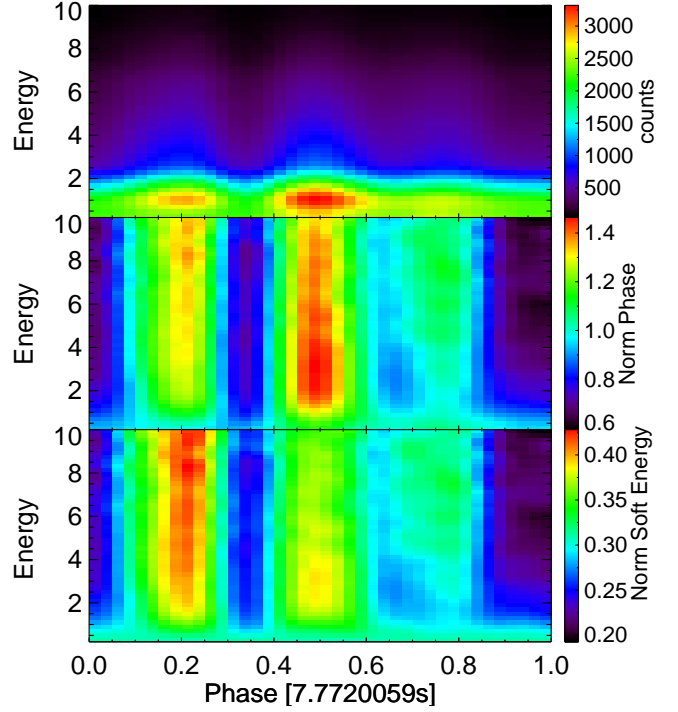}}
  \caption{Phase-dependent spectral heat map, obtained by binning the EPIC-pn source counts to 40 phase and 50 energy bins. 
           Top: Colors are proportional to the number of counts per bin.
           Middle: Data rates are corrected for the EPIC-pn response matrix efficiency and normalized by the phase-averaged number of counts in each energy bin.
           Bottom: Data normalized a second time using the rate average of each phase bin for the soft X-ray band (0.2-0.6 keV).  
  } 
  \label{fig:phase_heat}
\end{figure}

\subsection{X-ray spectral analysis}

We performed phase-averaged and phase-resolved spectroscopy of the pn data and a separate analysis of the RGS data, which was performed using the latest version of the {\small XSPEC} 
fitting package, Version 12.9.1 \citep{1996ASPC..101...17A}. 
The spectral continuum in both phase-averaged and phased-resolved spectroscopy was modeled using a combination of an absorbed multicolor disk blackbody (MCD) and a power-law component. The interstellar absorption was modeled using the {\texttt tbnew} code, which is a new and improved version of the X-ray absorption model {\texttt tbabs} \citep{2011_in_prep}. The atomic cross-sections were adopted
from \cite{1996ApJ...465..487V}. We considered a combination of Galactic foreground absorption and an additional column density accounting for both the interstellar medium
of the SMC and the intrinsic absorption of the source. For the Galactic photoelectric absorption, we considered a fixed column density of nH$_{\rm GAL}$ = 7.06$\times10^{20}$\,cm$^{−2}$ \citep{1990ARA&A..28..215D}, with abundances taken from \cite{2000ApJ...542..914W}.

\subsubsection{Phase-averaged spectroscopy}

Both thermal and nonthermal emission components were required to successfully fit the spectral continuum. More specifically, when modeling the continuum using a simple power law, there was a pronounced residual structure  that strongly indicated thermal emission below 1\,keV. On the other hand, nonthermal emission dominated the spectrum at higher energies. The thermal emission was modeled by an MCD model (for the rationale of this choice, see the discussion in Section~\ref{discussion}), which we modeled using the {\small XSPEC}  model {\texttt{diskbb}}. The continuum emission models alone could not fit the spectrum successfully, and strong emission-like residuals were noted at ${\sim}$0.51, ${\sim}$0.97, and ${\sim}$6.63\,keV and yielded a reduced $\chi^2$ value of 1.93 for 198 dof. (see ratio plot in Fig.~\ref{fig:spec_av}). The corresponding emission features where modeled using Gaussian curves,  yielding a $\chi^2$ value  of 1.01 for 190 dof. The three emission lines are centered at ${\rm E_{1}}{\sim}0.51$\,keV, ${\rm E_{2}}{\sim}0.96$\,keV, and ${\rm E_{3}}{\sim}6.64$\,keV and have equivalent width values of 6.6$_{-4.1}^{+5.0}$\,eV, 19$_{-5.1}^{+7.0}$\,eV, and 72$_{-15}^{+17}$\,eV, respectively. The ${\rm E_{2}}$ and ${\rm E_{3}}$ lines have a width of $83.8_{-2.6}^{3.0}$\,eV and 361$_{-80.1}^{+89.7}$\,eV, respectively, and the ${\rm E_{1}}$ line was not resolved.  The three emission-like features also appear in the two MOS detectors (see Fig.~\ref{fig:spec_av}) and the RGS data. The emission-like features in the MOS spectra increase the robustness of their detection, but the MOS data are not considered in this analysis. The MOS1 detector was heavily piled up, and since we used the pn detector for the phase-resolved spectroscopy (as it yields more than three times the number of counts of the MOS2 detector), we also opted to exclude the MOS2 data and only use the pn data for the phase-averaged spectroscopy in order to maintain consistency. For display purposes, we show the MOS1, MOS2, and pn spectra and the data-to-model vs energy plots in Fig.~\ref{fig:spec_av_MOS}. For the purposes of this plot, we fit the spectra of the three detectors using the {\texttt{diskbb}} plus {\texttt{powerlaw}} model, with all their parameters left free to vary. This choice allows for the visual detection of the line-like features while compensating for the considerable artificial hardening of the piled-up MOS1 detector. The MOS data indicate that the line-like features at ${\sim}0.5\,$keV and ${\sim}1\,$keV are composed of more emission lines that are unresolved by pn. This finding is confirmed by the RGS data (see Sect. \ref{sec:RGS} for more details). Narrow residual features are still present in the 1.7-2.5\,kev range. They stem most likely from incorrect modeling of the Si and Au absorption in the CCD detectors by the EPIC pn calibration, which often results in emission and/or absorption features at ${\sim}$1.84\,keV and ${\sim}$2.28\,keV (M$\beta$) and 2.4\,keV (M$\gamma$). Since the features are not too pronounced and did not affect the quality of the fit or the values of the best-fit parameters, they were ignored, but their respective energy channels were still included in our fits.

The best-fit parameters for the  modeling of the phase-averaged pn data are presented in Tables~\ref{tab:cont_phase} and~\ref{tab:lines}, with the normalization parameters of the two continuum components given in terms of their relevant fluxes (calculated using the multiplicative  {\small XSPEC}  model {\texttt{cflux}}). The phase-averaged pn spectrum along with the data-to-model ratio plot (without the Gaussian emission lines) is presented in Figure~\ref{fig:spec_av}. The phase-averaged spectrum was also modeled using a single-temperature blackbody instead of the MCD component. The fit was qualitatively similar to the MCD fit, with a k${\rm T_{BB}}$ temperature of 0.19$\pm0.01$\,keV, a size of 168$_{-16.4}^{+22.1}$\,km, and a power-law spectral index of $0.98\pm0.01$.

\begin{figure}
         \resizebox{\hsize}{!}{\includegraphics[angle=-90,clip=]{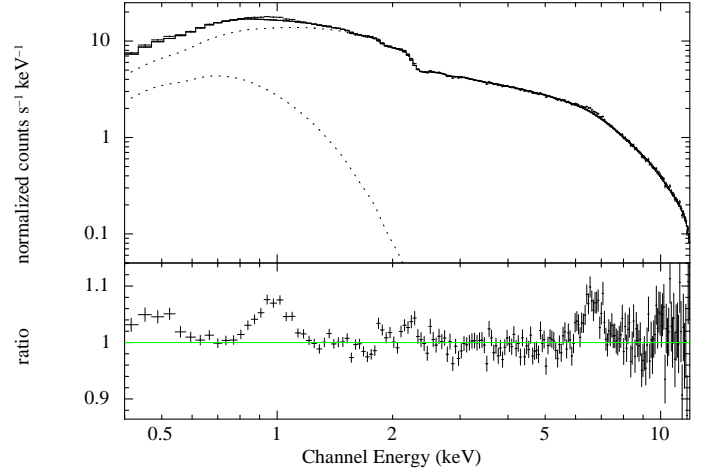}}
  \caption{Phase-averaged X-ray spectrum without the Gaussian emission lines. { Top plot:} Normalized counts vs energy. { Bottom plot:} Data-to-model ratio plot. }   
  \label{fig:spec_av}
\end{figure}

\begin{figure}
         \resizebox{\hsize}{!}{\includegraphics[angle=-90,clip=]{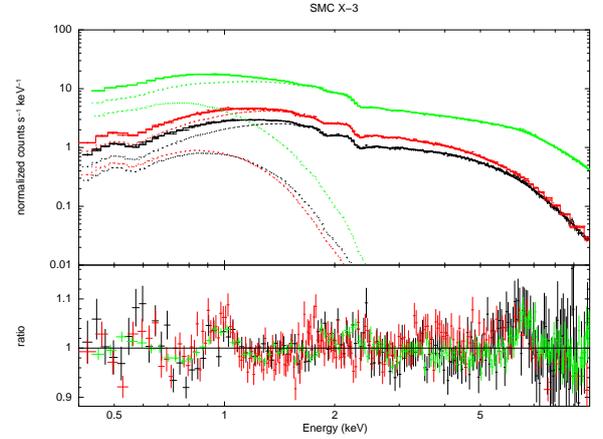}}
  \caption{Phase-averaged X-ray spectrum with only the continuum model for all EPIC detectors (black: MOS1, red: MOS2, green: pn). { Top plot:} Normalized counts vs energy. { Bottom plot:} Data-to-model ratio plot. }   
  \label{fig:spec_av_MOS}
\end{figure}

\subsubsection{Phase-resolved spectroscopy}
\label{phase-res}

We created separate spectra from 20 phase intervals of equal duration. We studied the resulting phase-resolved spectra by fitting each individual spectrum separately using the same model as in the phase-averaged spectrum. We noted the behavior of the two spectral components and the three emission lines, and monitored the variation of their best-fit parameters as the pulse phase evolved. The resulting best-fit values for the continuum are presented in Table~\ref{tab:cont_phase}, and those of the emission lines in Table~\ref{tab:lines}. In Figure~\ref{fig:phace_spec_res} we present the evolution of the different spectral parameters and the source count rate with the pulse phase.  

For purposes of presentation and to further note the persistent nature of the soft emission, we also fit all 20 spectra simultaneously and tied the parameters of the thermal component, while the parameters of the power-law component were left free to vary. The results of this analysis were qualitatively similar to the independent spectral fitting (see Fig.~\ref{fig:spec_res}, where the spectral evolution of the source is illustrated). Nevertheless, the decision to freeze the temperature and normalization of the soft component at the phase-averaged best-fit value introduces bias to our fit, and we therefore only tabulate and study the best-fit values for the independent modeling of the phase-resolved spectra.

\begin{figure}
   \resizebox{\hsize}{!}{\includegraphics[angle=-90,clip=]{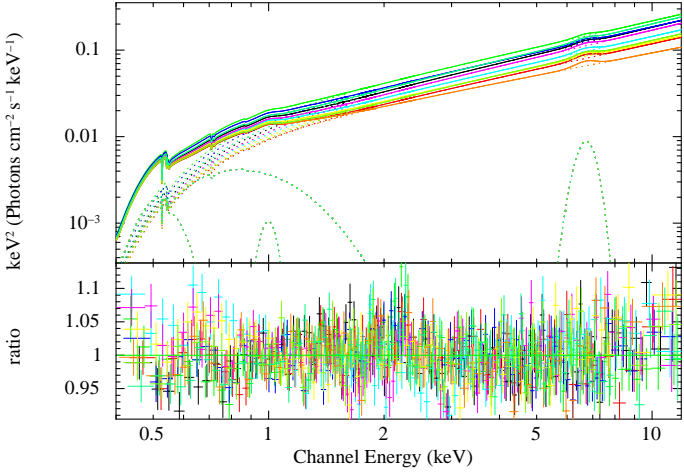}}
  \caption{Spectral model and data-to-ratio plot of the simultaneous fit of the 20 phase-resolved spectra }   
  \label{fig:spec_res}
\end{figure}

\begin{figure}
\resizebox{\hsize}{!}{\includegraphics[angle=0,clip=, bb= 0 24 420 710]{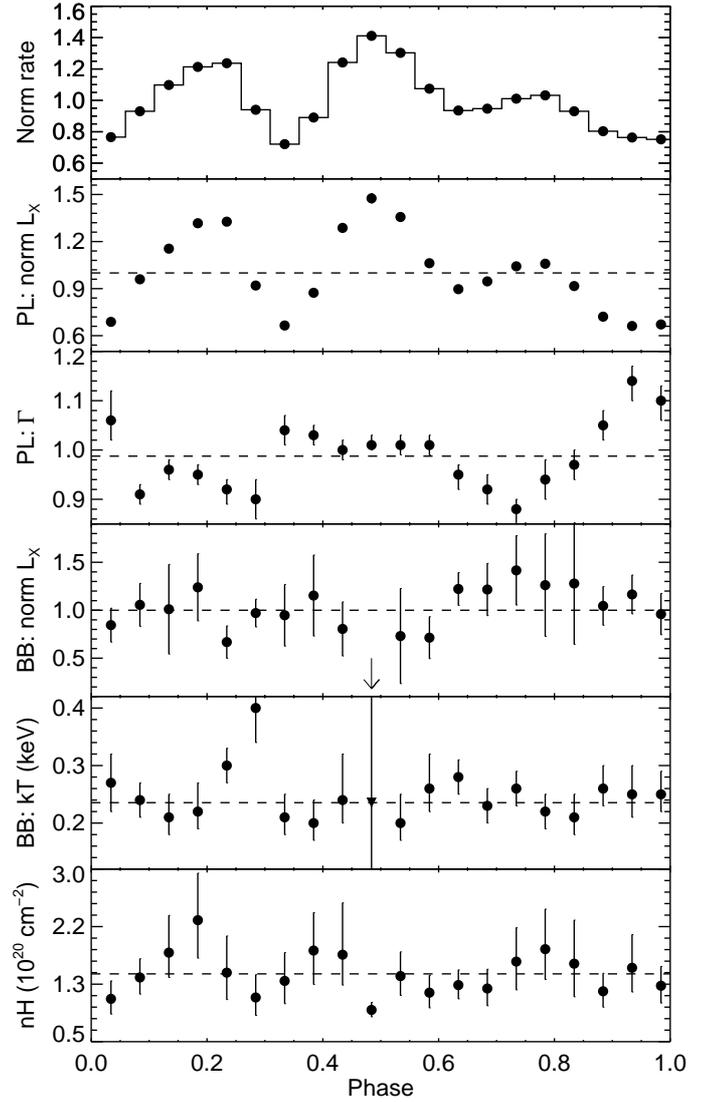}}
  \caption{Evolution of spectral parameters (Table~\ref{tab:cont_phase}) for the phase-resolved spectroscopy presented in Sect. \ref{phase-res}. The arrow represents the upper limit for the flux of the MCD component for phase-resolved spectrum 5. The corresponding temperature was frozen to the phase-averaged value (see the first row of Table~\ref{tab:cont_phase}).}   
  \label{fig:phace_spec_res}
\end{figure}

\subsubsection{RGS spectroscopy}
\label{sec:RGS}

The RGS spectra  were not grouped and were fitted employing Cash statistics \citep{1979ApJ...228..939C} and with the same continuum model as the phased-averaged pn spectra.
The model yielded an acceptable fit, with a reduced $\chi^2$ value of 1.05 for 1875 dof.  The RGS spectra were also fitted simultaneously with the pn phase-averaged spectra (with the addition of a normalization constant, which was left free to vary) yielding best-fit parameter values consistent within 1\,${\sigma}$ error bars with the pn continuum model. We therefore froze the continuum parameters to the values tabulated in the first row of Table~\ref{tab:cont_phase}. We note that a simple absorbed power-law model still describes the continuum with sufficient accuracy, and the use of a second component only improves the fit by $\sim$2.7$\sigma,$ as indicated by performing an ftest\footnote{Algorithms taken from the CEPHES special function library by Stephen; Moshier. (\url{http://www.netlib.org/cephes/})}. 

We searched the RGS spectra for emission and absorption lines by a blind search for Gaussian features with a fixed width.  The spectrum was searched by adding the Gaussian to the continuum and fitting the spectrum with a step of 10\,eV, and the resulting $\Delta$C was estimated for each step.  Emission and absorption lines are detected with a significance higher than 3${\sigma.}$   To estimate  the values for the $\Delta$C that correspond to the 3$\sigma$ significance, we followed a similar approach as \cite{2018MNRAS.473.5680K}. We simulated 1000 spectra based on the continuum model and repeated the process of line-search using the fixed-width Gaussian.  We plot the probability distribution of the $\Delta$C values and indicate the range of $\Delta$C values within which lies the  99.7\% of the trials. $\Delta$C values outside this interval (see Fig.~\ref{fig:spec_RGS}, dotted lines) are defined as having a 3$\sigma$ significance. In Fig.~\ref{fig:spec_RGS} we plot the $\Delta$C improvement of emission and absorption lines detected in the RGS data assuming Gaussian lines with a fixed width of 2\,eV (black solid line) and 10\,eV (orange dashed line). The line significance is affected by the width of the Gaussian, and a more thorough approach should include a grid of different line widths in order to achieve the best fit. We here used the emission lines with a significance
higher  than 3\,$\sigma$  (dotted line) assuming a 2\,eV fixed width, which were then fit manually with the width as a free parameter. The best-fit values are presented in Table \ref{tab:rgs}. We also note a strong absorption line at more than 5\,$\sigma$ significance  and several tentative others. The most prominent absorption line is centered at ${\approx}1.585\,$keV. When the line is associated with MgXII absorption, it appears to be blueshifted by 7\%. The apparent blueshift could be an indication of outflows, but he velocity appears to be considerable lower than the velocities inferred for ULXs \citep[e.g.,][]{2016Natur.533...64P}. A more thorough study of the RGS data using an extended grid of all line parameters will be part of a separate publication.

\begin{table*}[!htbp]
 \caption{Best-fit parameters for the continuum for the 20 phases and the phase-averaged spectrum. All errors are in the 90\% confidence range.}
 \begin{center}
\scalebox{0.8}{   \begin{tabular}{lccccccccc}
     \hline\hline\noalign{\smallskip}
     \multicolumn{1}{c}{Phase} &
     \multicolumn{1}{c}{nH $^{[a]}$} &
     \multicolumn{1}{c}{k${\rm T_{\rm soft}}$} &
     \multicolumn{1}{c}{ F$_{soft}$ $^{[b]}$} &
     \multicolumn{1}{c}{${\Gamma}$} &
     \multicolumn{1}{c}{F$_{\rm PO}$ $^{[c]}$} &
     \multicolumn{1}{c}{count-rate} &
     \multicolumn{1}{r}{$\chi^2/dof$ }\\
     \noalign{\smallskip}\hline\noalign{\smallskip}
      
      \multicolumn{1}{c}{} &
      \multicolumn{1}{c}{[$\times10^{21}$\,cm$^2$]} &     
      \multicolumn{1}{c}{keV} &
      \multicolumn{1}{c}{$10^{-11}{\rm erg\,cm^{-2}\,s^{-1}}$} &
      \multicolumn{1}{c}{} &
      \multicolumn{1}{c}{$10^{-11}{\rm erg\,cm^{-2}\,s^{-1}}$} &
      \multicolumn{1}{c}{cts/s}  &
      \multicolumn{1}{c}{} \\
      \noalign{\smallskip}\hline\noalign{\smallskip}

      \noalign{\smallskip}\hline\noalign{\smallskip}      
    Average &    1.36$_{-0.62}^{+0.13}$     &0.24$_{-0.03}^{+0.04}$  &  1.44${\pm 0.19}$ & 0.98${\pm 0.10}$       &  29.8$\pm0.95$     & 41.8$\pm0.04$ & 1.01/190  \\
      \noalign{\smallskip}\hline\noalign{\smallskip}
       0    &    1.50$_{-0.39}^{+0.53}$     &  0.30$\pm0.03$         &  1.17${\pm 0.30}$&  0.92$_{-0.03}^{+0.02}$ &  40.1${\pm 0.37}$  & 52.5$\pm0.19$ & 1.05/181  \\
       1    &    1.14$_{-0.26}^{+0.33}$     &  0.40$_{-0.06}^{+0.07}$&  1.70${\pm 0.25}$&  0.90$\pm0.04$          &  27.8${\pm 0.29}$  & 39.9$\pm0.17$ & 0.95/177  \\
       2    &    1.38$_{-0.33}^{+0.41}$     &  0.21$_{-0.03}^{+0.04}$&  1.66${\pm 0.60}$&  1.04$\pm0.03$          &  20.1${\pm 0.25}$  & 30.6$\pm0.17$ & 1.01/169  \\
       3    &    1.82$_{-0.49}^{+0.55}$     &  0.20$_{-0.03}^{+0.04}$&  2.02${\pm 0.74}$&  1.03$\pm0.02$          &  26.4${\pm 0.28}$  & 37.8$\pm0.17$ & 1.03/171  \\
       4    &    1.76$_{-0.44}^{+0.75}$     &  0.24$_{-0.04}^{+0.08}$&  1.41${\pm 0.50}$&  1.00$\pm0.02$          &  38.9${\pm 0.35}$  & 52.7$\pm0.21$ & 1.00/183  \\
       5    &    0.96$_{-0.10}^{+0.11}$     &  --                    &  $<$2.17         &  1.01$\pm0.01$          &  44.6${\pm 0.90}$  & 59.9$\pm0.20$ & 0.87/184  \\
       6    &    1.45$_{-0.28}^{+0.35}$     &  0.20$_{-0.03}^{+0.05}$&  1.28${\pm 0.87}$&  1.01$\pm0.02$          &  41.0${\pm 0.35}$  & 55.3$\pm0.19$ & 1.01/187  \\
       7    &    1.21$_{-0.22}^{+0.25}$     &  0.26$_{-0.04}^{+0.06}$&  1.25${\pm 0.38}$&  1.01$\pm0.02$          &  32.1${\pm 0.37}$  & 45.6$\pm0.20$ & 0.89/182  \\
       8    &    1.32$_{-0.20}^{+0.22}$     &  0.28$\pm0.03$         &  2.14${\pm 0.30}$&  0.95$\pm0.03$          &  27.1${\pm 0.34}$  & 39.7$\pm0.17$ & 0.81/178  \\
       9    &    1.27$_{-0.25}^{+0.28}$     &  0.23$\pm0.03$         &  2.13${\pm 0.48}$&  0.92$\pm0.03$          &  28.6${\pm 0.40}$  & 40.2$\pm0.17$ & 0.88/178  \\
       10   &    1.66$_{-0.41}^{+0.49}$     &  0.26$\pm0.03$         &  2.48${\pm 0.63}$&  0.88$_{-0.03}^{+0.02}$ &  31.5${\pm 0.28}$  & 42.9$\pm0.18$ & 0.87/182  \\
       11   &    1.84$_{-0.44}^{+0.58}$     &  0.22$\pm0.03$         &  2.21${\pm 0.94}$&  0.94$\pm0.04$          &  32.0${\pm 0.46}$  & 43.8$\pm0.18$ & 1.08/178  \\
       12   &    1.63$_{-0.48}^{+0.63}$     &  0.21$_{-0.03}^{+0.04}$&  2.24${\pm 1.11}$&  0.97$\pm0.03$          &  27.7${\pm 0.37}$  & 39.5$\pm0.17$ & 1.09/173  \\
       13   &    1.23$_{-0.23}^{+0.26}$     &  0.26$_{-0.03}^{+0.04}$&  1.83${\pm 0.35}$&  1.05$\pm0.03$          &  21.8${\pm 0.26}$  & 34.1$\pm0.17$ & 0.99/173  \\
       14   &    1.57$_{-0.35}^{+0.48}$     &  0.25$_{-0.04}^{+0.05}$&  2.04${\pm 0.35}$&  1.14$_{-0.04}^{+0.03}$ &  20.0${\pm 0.25}$  & 32.4$\pm0.15$ & 0.89/170  \\
       15   &    1.31$_{-0.25}^{+0.28}$     &  0.25$_{-0.03}^{+0.04}$&  1.68${\pm 0.38}$&  1.10$_{-0.04}^{+0.03}$ &  20.3${\pm 0.25}$  & 31.9$\pm0.15$ & 1.07/170  \\
       16   &    1.12$_{-0.22}^{+0.26}$     &  0.27$\pm0.05$         &  0.15${\pm 0.31}$&  1.06$_{-0.04}^{+0.06}$ &  20.9${\pm 0.26}$  & 32.5$\pm0.15$ & 1.01/172  \\
       17   &    1.43$_{-0.24}^{+0.27}$     &  0.24$\pm0.03$         &  1.85${\pm 0.39}$&  0.91$\pm0.02$          &  29.0${\pm 0.31}$  & 39.5$\pm0.17$ & 0.94/187  \\
       18   &    1.79$_{-0.36}^{+0.54}$     &  0.21$_{-0.03}^{+0.04}$&  1.77${\pm 0.82}$&  0.96$\pm0.02$          &  34.9${\pm 0.33}$  & 46.6$\pm0.18$ & 1.11/178  \\
       19   &    2.26$_{-0.55}^{+0.68}$     &  0.22$_{-0.03}^{+0.05}$&  2.17${\pm 0.61}$&  0.95$\pm0.02$          &  39.8${\pm 0.36}$  & 51.5$\pm0.19$ & 0.96/182  \\

      \noalign{\smallskip}\hline\noalign{\smallskip}         
    \end{tabular}   }
 \end{center}
  \tablefoot{
  \tablefoottext{a}{Column density of the intrinsic absorption. The Galactic foreground absorption was fixed to a value of nH$_{\rm GAL}$ = 7.06$\times10^{20}$\,cm$^{−2}$ \citep{1990ARA&A..28..215D}, with abundances taken from \cite{2000ApJ...542..914W}.}\\
  \tablefoottext{b}{Unabsorbed bolometric X-ray flux of the BB in the 0.2-12\,keV band.}\\
\tablefoottext{c}{Unabsorbed X-ray flux in the 0.2-12\,keV band. For the distance of 62\,kpc \citep{2014ApJ...780...59G}, a conversion factor of $4.6\times10^{47} cm^{-2}$ can be used to convert from flux into luminosity.}\\
  }
 \label{tab:cont_phase}
\end{table*}

\begin{table*}[!htbp]
 \caption{Bestfit parameters for the emission lines for the 20 phases. All errors are in the 90\% confidence range.}
 \begin{center}
\scalebox{0.8}{   \begin{tabular}{lccccccccccc}
     \hline\hline\noalign{\smallskip}
     \multicolumn{1}{c}{Phase} &
     \multicolumn{1}{c}{${\rm E_{1}}$} &
     \multicolumn{1}{c}{${\rm K_{1}}$} &
     \multicolumn{1}{c}{${\rm E_{2}}$} &
     \multicolumn{1}{c}{${\rm K_{3}}$} &
     \multicolumn{1}{c}{${\rm E_{3}}$} &
     \multicolumn{1}{c}{${\rm {\sigma}_{3}}$} &
     \multicolumn{1}{c}{${\rm K_{3}}$} \\
     \noalign{\smallskip}\hline\noalign{\smallskip}
      
      \multicolumn{1}{c}{} &
      \multicolumn{1}{c}{keV} &
      \multicolumn{1}{c}{$10^{-4}$\,cts\,cm$^{-2}$\,s$^{-1}$} &
      \multicolumn{1}{c}{keV} &
      \multicolumn{1}{c}{$10^{-4}$\,cts\,cm$^{-2}$\,s$^{-1}$} &
      \multicolumn{1}{c}{keV} &
      \multicolumn{1}{c}{keV} &
      \multicolumn{1}{c}{$10^{-4}$\,cts\,cm$^{-2}$\,s$^{-1}$}  \\
      \noalign{\smallskip}\hline\noalign{\smallskip}

      \noalign{\smallskip}\hline\noalign{\smallskip}
    Average$^{a}$ & 0.51$\pm0.02$          & 3.02$_{-1.78}^{+2.36}$ & 0.96$\pm0.02$         & 3.55$_{-0.98}^{+1.32}$ &  6.64$_{-0.15}^{+0.16}$ & 0.36$_{-0.08}^{+0.09}$ &  1.70$_{-0.35}^{+0.39}$  \\
      \noalign{\smallskip}\hline\noalign{\smallskip}
       0    & 0.49$\pm0.03$          & 11.2$_{-8.52}^{+15.8}$ & 0.98$\pm0.03$         & 2.79$_{-1.19}^{+1.18}$ &  6.76$_{-0.15}^{+0.16}$ & 0.25$_{-0.15}^{+0.20}$ &  2.43$_{-1.23}^{+1.40}$  \\
       1    & 0.51$_{-0.03}^{+0.04}$ & 6.98$_{-5.27}^{+8.06}$ & 0.94$_{-0.05}^{+0.06}$& 1.91$\pm{1.13}$        &  6.44$_{-0.32}^{+0.20}$ & $<0.39$                &  1.33$_{-0.82}^{+0.96}$  \\
       2    & 0.53$_{-0.15}^{+0.09}$ &    $<$10.8             & 1.02$_{-0.03}^{+0.02}$& 2.69$_{-0.89}^{+0.97}$ &  6.76$_{-0.30}^{+0.26}$ & 0.45*                  &  2.28$_{-1.15}^{+1.10}$  \\
       3    & 0.50$_{-0.05}^{+0.06}$ &    $<$21.9             & 0.97$\pm0.03$         & 2.75$_{-1.16}^{+1.19}$ &  6.49$_{-0.26}^{+0.14}$ & $<0.48$                &  0.92$_{-0.62}^{+1.13}$  \\
       4    & 0.48$_{-0.02}^{+0.03}$ & 16.8$_{-12.5}^{+23.5}$ & --                    & --                     &  --                     &  --                    &  --                      \\
       5    & 0.55$_{-0.04}^{+0.08}$ & 4.96$_{-4.31}^{+5.21}$ & 0.98$\pm0.03$         & 3.12$\pm{1.13}$        &  6.61$_{-0.30}^{+0.26}$ & 0.41$_{-0.22}^{+0.34}$ &  3.06$_{-1.53}^{+2.08}$  \\
       6    & --                     & --                     & --                    & --                     &  --                     &  --                    &  --                      \\
       7    & --                     & --                     & 0.94$\pm0.03$         & 2.69$\pm{1.23}$        &  --                     & --                     & --                       \\
       8    & --                     & --                     & --                    & --                     &  6.74$_{-0.24}^{+0.22}$ & 0.28$_{-0.24}^{+0.38}$ &  1.80$_{-1.07}^{+1.54}$  \\
       9    & --                     & --                     & 0.98$_{-0.02}^{+0.03}$& 3.09$_{-1.08}^{+1.10}$ &  6.81$_{-0.28}^{+0.31}$ & 0.45*                  &  2.93$\pm{1.13}$         \\
       10   & 0.46$\pm0.04$          & 16.0$_{-13.1}^{+21.9}$ & --                    & --                     &  --                     &  --                    &  --                      \\
       11   & 0.50$_{-0.02}^{+0.03}$ & 14.7$_{-10.6}^{+20.2}$ & 0.94$\pm0.03$         & 3.12$\pm{1.17}$        &  6.76$_{-0.39}^{+0.29}$ & 0.45*                  &  2.97$_{-1.30}^{+1.29}$  \\
       12   & 0.46$_{-0.04}^{+0.03}$ & $<$8.90                & 0.97$\pm0.03$         & 2.30$_{-1.13}^{+1.14}$ &  6.70$_{-0.19}^{+0.20}$ & 0.45*                  &  3.60$_{-1.41}^{+1.25}$  \\
       13   & --                     & --                     & 0.96$_{-0.09}^{+0.05}$& 1.70$_{-1.06}^{+1.07}$ &  --                     & --                     &  --                      \\
       14   & 0.51$_{-0.05}^{+0.08}$ & 7.49$_{-6.58}^{+14.1}$ & 0.99$_{-0.04}^{+0.03}$& 1.85$_{-1.03}^{+1.04}$ &  6.9*                   & 0.44$_{-0.16}^{+0.20}$ &  2.52$_{-1.14}^{+1.25}$  \\
       15   & --                     & --                     & 0.95$_{-0.07}^{+0.16}$& 1.13$_{-1.04}^{+1.06}$ &  6.87$_{-0.15}^{+0.19}$ & 0.30$_{-0.12}^{+0.22}$ &  2.32$_{-0.93}^{+1.04}$  \\
       16   & --                     & --                     & 1.02$_{-0.08}^{+0.04}$& 1.76$_{-0.93}^{+0.94}$ &  6.42$_{-0.19}^{+0.42}$ & 0.30$_{-0.14}^{+0.39}$ &  1.79$_{-0.90}^{+1.50}$  \\
       17   & --                     & --                     & --                    & --                     &  6.58$_{-0.11}^{+0.12}$ & $<0.29$                &  1.67$_{-0.87}^{+0.90}$  \\
       18   & 0.53$\pm0.03$          & 6.41$_{-5.31}^{+5.44}$ & 1.00$_{-0.04}^{+0.03}$& 2.25$_{-1.17}^{+1.12}$ &  6.45$\pm0.06$          & $<0.14$                &  1.06$\pm{0.53}$         \\
       19   & 0.49$\pm0.02$          & 24.2$_{-16.0}^{+29.6}$ & --                    & --                     &  6.60$_{-0.36}^{+0.13}$ & $<0.60$                &  1.38$_{-0.91}^{+1.88}$  \\

      \noalign{\smallskip}\hline\noalign{\smallskip}         

    \end{tabular}   }
 \end{center}
  \tablefoot{
  \tablefoottext{a}{The line widths E$_{1}$ and E$_{2}$ were frozen to the value of 1\,ev in the phase-resolved analysis because they are too narrow to be resolved by the pn detector. Nevertheless, E$_{2}$ in the phase-averaged spectra is broader and has a width value of $\sigma=83.8_{-2.56}^{3.04}$\,eV.}\\
   \tablefoottext{*}{Parameter frozen.}}
 \label{tab:lines}
\end{table*}

\begin{table}[htbp]
 \caption{RGS emission lines and the corresponding the $\Delta$C value for each added Gaussian. All errors are in the 90\% confidence range.}
\scalebox{0.8}{   \begin{tabular}{lcccccccc}
     \hline\hline\noalign{\smallskip}
     \multicolumn{1}{c}{Line$^{a}$} &
     \multicolumn{1}{c}{${\rm E}$} &
     \multicolumn{1}{c}{${\rm {\sigma}}$} &
     \multicolumn{1}{c}{${\rm norm}$} &
     \multicolumn{1}{c}{{\rm ${\Delta}$C}} \\

     \noalign{\smallskip}\hline\noalign{\smallskip}
      
      \multicolumn{1}{c}{eV} &
      \multicolumn{1}{c}{keV} &
      \multicolumn{1}{c}{eV} &
      \multicolumn{1}{c}{$10^{-4}$\,cts\,cm$^{-2}$\,s$^{-1}$} \\
      \noalign{\smallskip}\hline\noalign{\smallskip}

      \noalign{\smallskip}\hline\noalign{\smallskip}
        \ion{N}{VII} (500.3)      & 0.512$_{-0.014}^{+0.011}$ & 16.02$_{-6.44}^{+13.00}$ & 3.152$_{-1.506}^{+1.510}$ & 12.5\\
        \ion{O}{VII} (561-574)    & 0.571$\pm0.005          $ & $<10.01$                 & 1.341$_{-0.893}^{+0.901}$ & 7.29\\
        \ion{O}{VIII}  (653.6)    & 0.657$_{-0.03}^{+0.003}$  & $<7.0$                   & 0.6$_{-0.3}^{+0.5}$       & 14.9   \\
       \ion{Fe}{XIX-XXI}(917.2)   & 0.923$\pm0.006          $ & 10.40$_{-3.39}^{+6.73}$  & 0.964$_{-0.371}^{+0.377}$ & 20.3\\
       \ion{Ne}{X}  (1022)        & 1.025$_{-0.002}^{+0.003}$ & 3.481$_{-2.017}^{+15.09}$& 0.653$_{-0.246}^{+0.247}$ & 20.6\\
       \ion{Fe}{XXIII-XXIV} (1125)& 1.125$_{-0.020}^{+0.018}$ & 16.77$_{-11.55}^{+24.61}$& 0.597$\pm0.413          $ & 5.8\\
                                                           
      \noalign{\smallskip}\hline\noalign{\smallskip}         

    \end{tabular}   }
  \tablefoot{
  \tablefoottext{a}{Most likely emission lines assuming collisionally
ionized diffuse gas at a temperature of ${\sim}1$\,keV. }\\
  }
 \label{tab:rgs}
\end{table}

\begin{figure*}
                  \resizebox{\hsize}{!}{\includegraphics[angle=0]{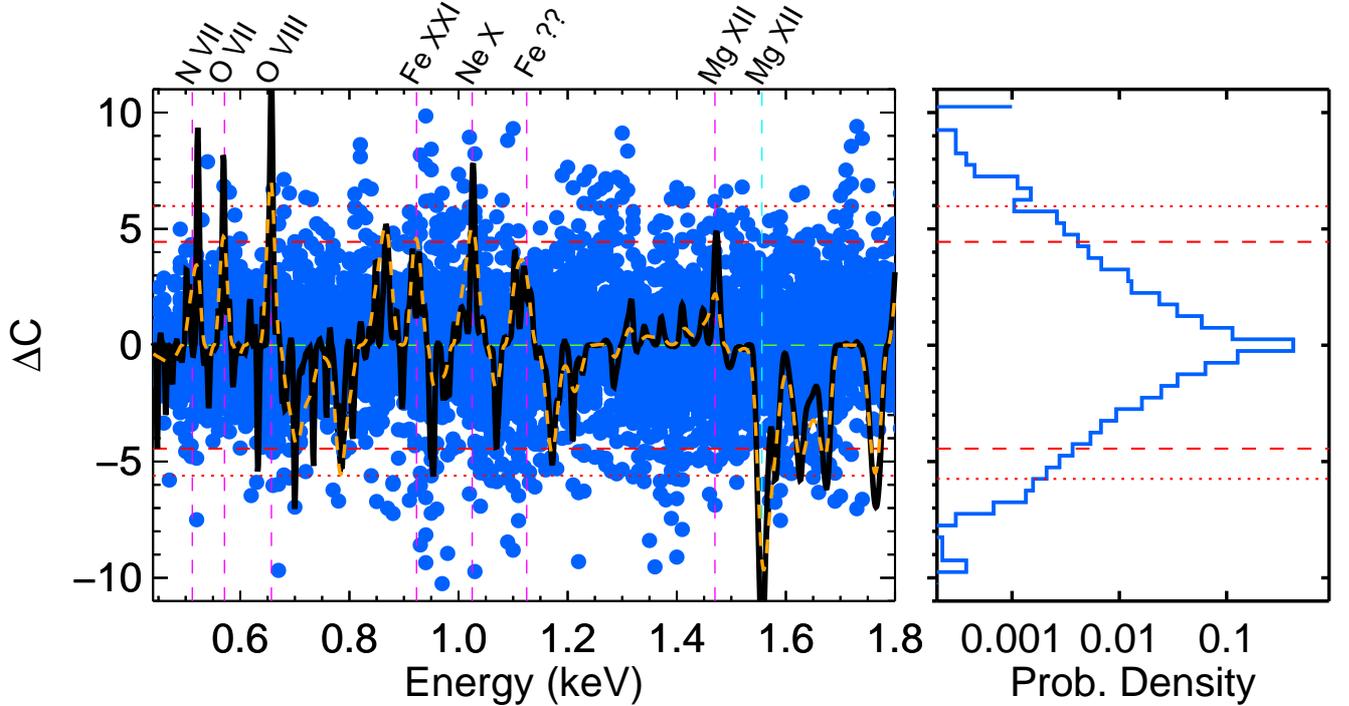}}
  \caption{Emission and absorption lines detected in the RGS data. { Left plot:} ${\Delta}$C for each line vs energy. Black corresponds to Gaussians with 2\,eV width and orange to a width
of 10\,eV. The blue points correspond to the simulated spectra of the model continuum. { Right plot:} Probability density function for ${\Delta}$C values for the simulated spectra.  The
dashed line corresponds to three standard deviations assuming that the probability density function follows the Gaussian distribution. The dotted line encloses 99.7\% of the  ${\Delta}$C  values. }   
  \label{fig:spec_RGS}
\end{figure*}

\subsection{X-ray spectral and temporal evolution during the 2016 outburst}

We report on the long-term evolution of the values of the PF and HR throughout the 2016 outburst, using the data from 
\swift observatory.
Using the  method described in Sect. \ref{time}, we searched for the pulse period of \sxp for all the analyzed \swift/XRT observations 
using the barycenter corrected events within the 0.5-10.0 keV energy band. The resulting pulse periods are presented in Fig. \ref{fig:t_spin}.
Using the derived best-fit periods, we estimated  the PF for all the \swift/XRT observations (see Fig. \ref{fig:cr_pf}).
From the extracted event files, we calculated the phase-average hardness ratios for all \swift/XRT observations.
For simplicity and in order to increase statistics, we used three energy bands to estimate two HR values: HR { soft,} using the 0.5-2.0 keV and 2.0-4.5 keV bands, and HR { hard,} using the 2.0-4.5 keV and 4.5-10.0 keV bands. 

We performed the same calculations for the phase-resolved spectrum of \sxp as measured by the \xmm ToO. 
In particular, we estimated the above two HR values for 40 phase intervals.
The evolution of the spectral state of \sxp with time and pulse phase as depicted by these two HR values is shown in Fig.~\ref{fig:HR_L} (left). 
For comparison, we have plot in parallel (Fig.~\ref{fig:HR_L}, right panel) the evolution of the HR with NS spin phase during the \xmm observation. We note that although quantitatively different  HR values are expected for different instruments (i.e., \swift/XRT and \xmm/EPIC-pn),  a qualitative comparison of the HR evolution can be drawn. More specifically, the left-hand plot  probes the long-term HR evolution versus the mass accretion rate (inferred from the registered source count-rate), and the right-hand side shows the HR evolution with pulse phase.

\begin{figure}
   \resizebox{\hsize}{!}{\includegraphics[angle=0,clip=, bb= 0 31 510 500]{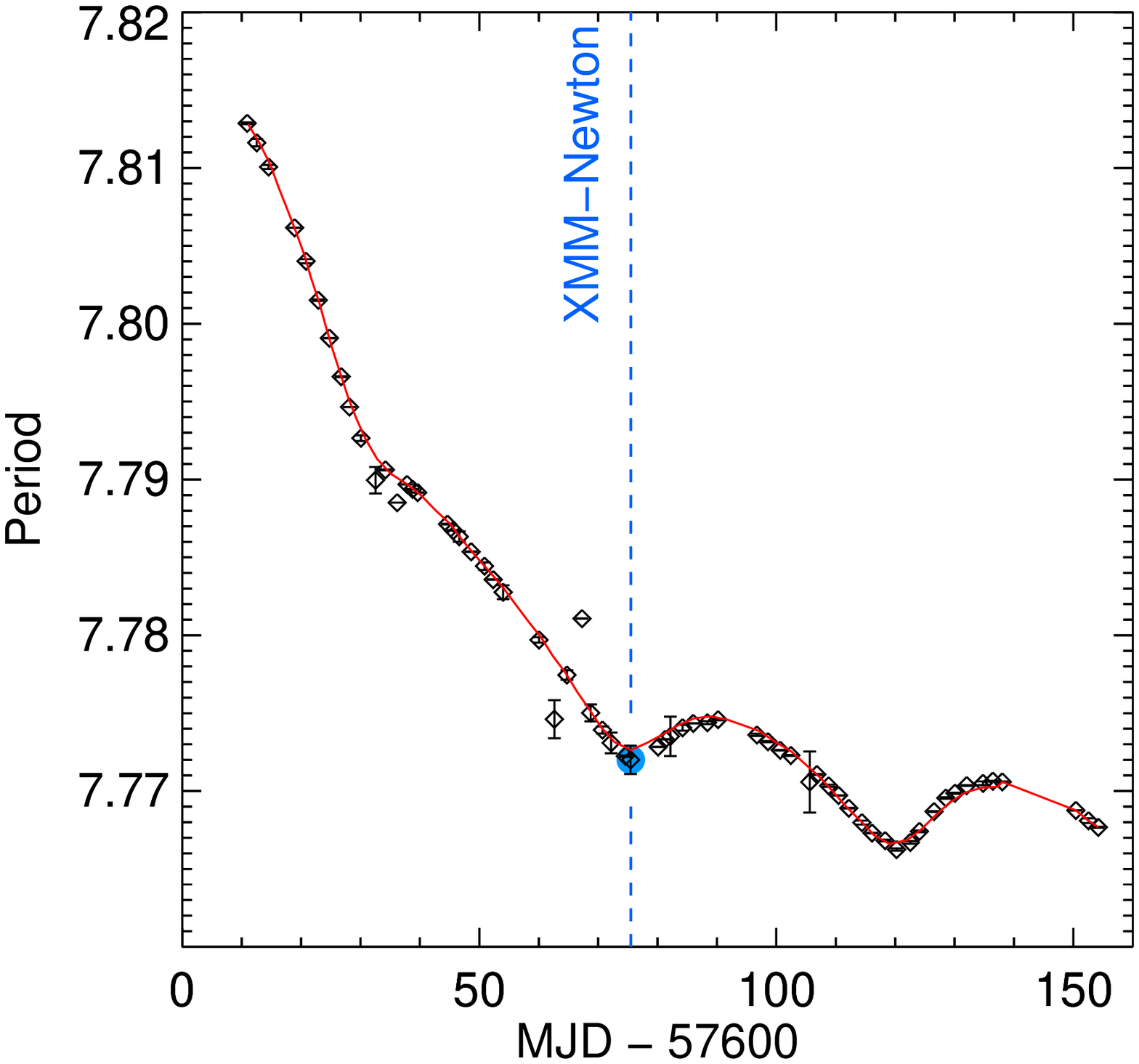}}
  \caption{Pulse period evolution during the 2016 burst of \sxp based on the analysis of \swift/XRT data. The red line is based on the best-fit model describing the evolution of the spin \citep{2017arXiv170102336T}. The date of the \xmm pointing is marked by the blue dashed vertical line.
  The spin period derived by the EPIC-pn data is marked with a blue circle. This solution coincides with the one derived from \swift/XRT on an observation performed within the same day.}   
  \label{fig:t_spin}
\end{figure}

\begin{figure}
   \resizebox{\hsize}{!}{\includegraphics[angle=0,clip=, bb= 0 15 500 400]{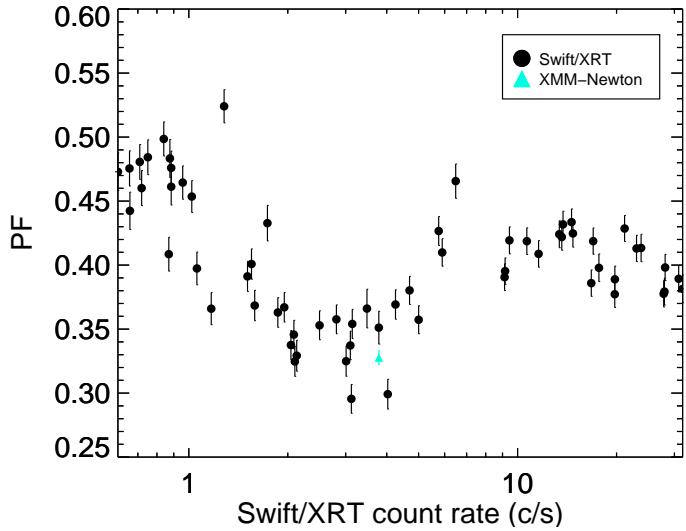}}
   \caption{PF: Pulsed fraction defined in equation \ref{pfeq}. \swift errors are computed based on the Poisson uncertainty of the pulse profile and do not include calculations based on the period uncertainties, as a systematic uncertainty of 0.01 was added to the estimated \swift/XRT PF. The \xmm PF is estimated from the profile presented in Fig. \ref{fig:pp}, and errors are estimated from the 1$\sigma$ error in the derived period. }
   \label{fig:cr_pf}
\end{figure}  

\begin{figure*}
   \resizebox{\hsize}{!}{\includegraphics[angle=0,clip=, bb= 33 38 519 522]{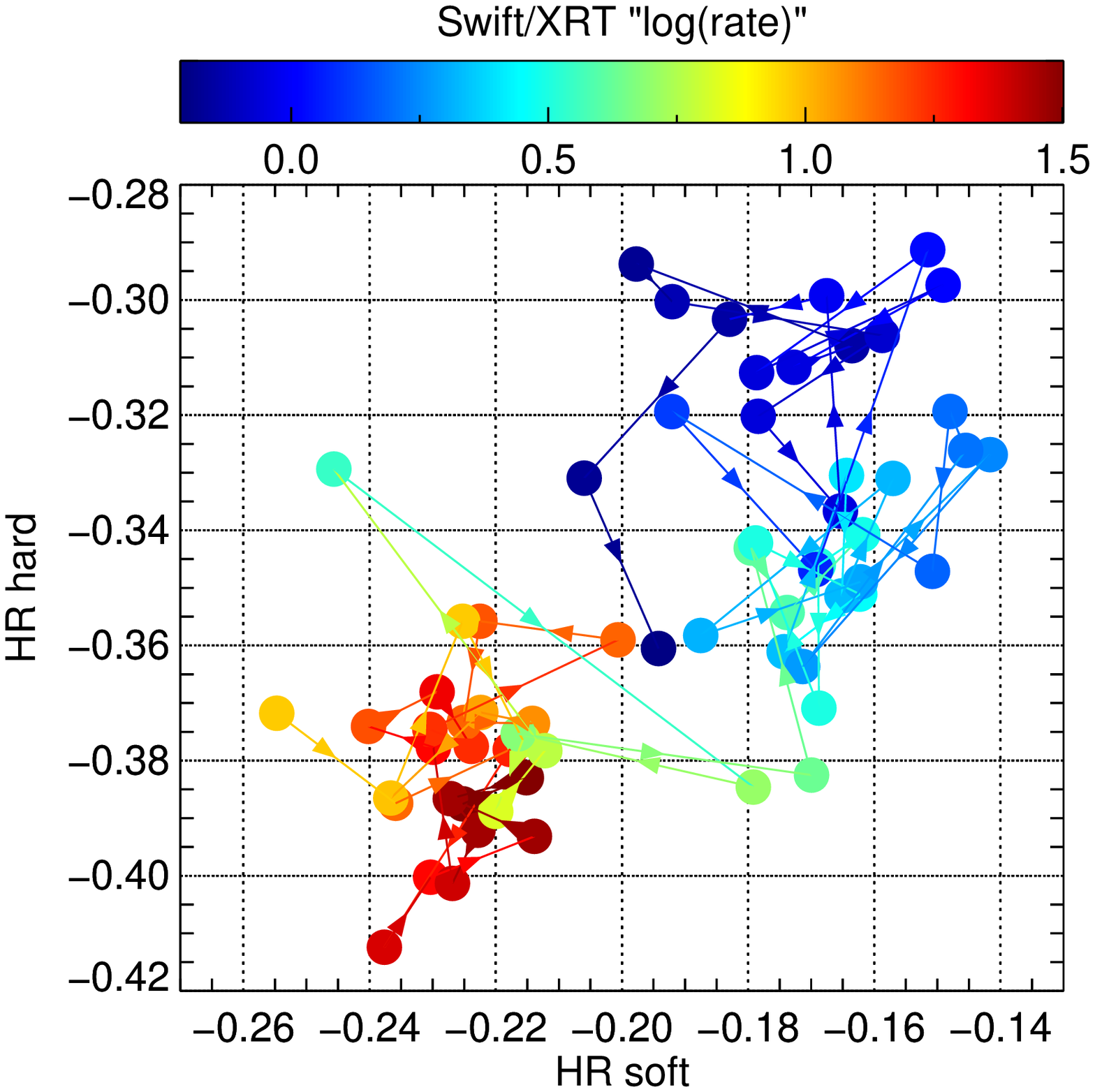}
\includegraphics[angle=0,clip=, bb= 33 38 519 522]{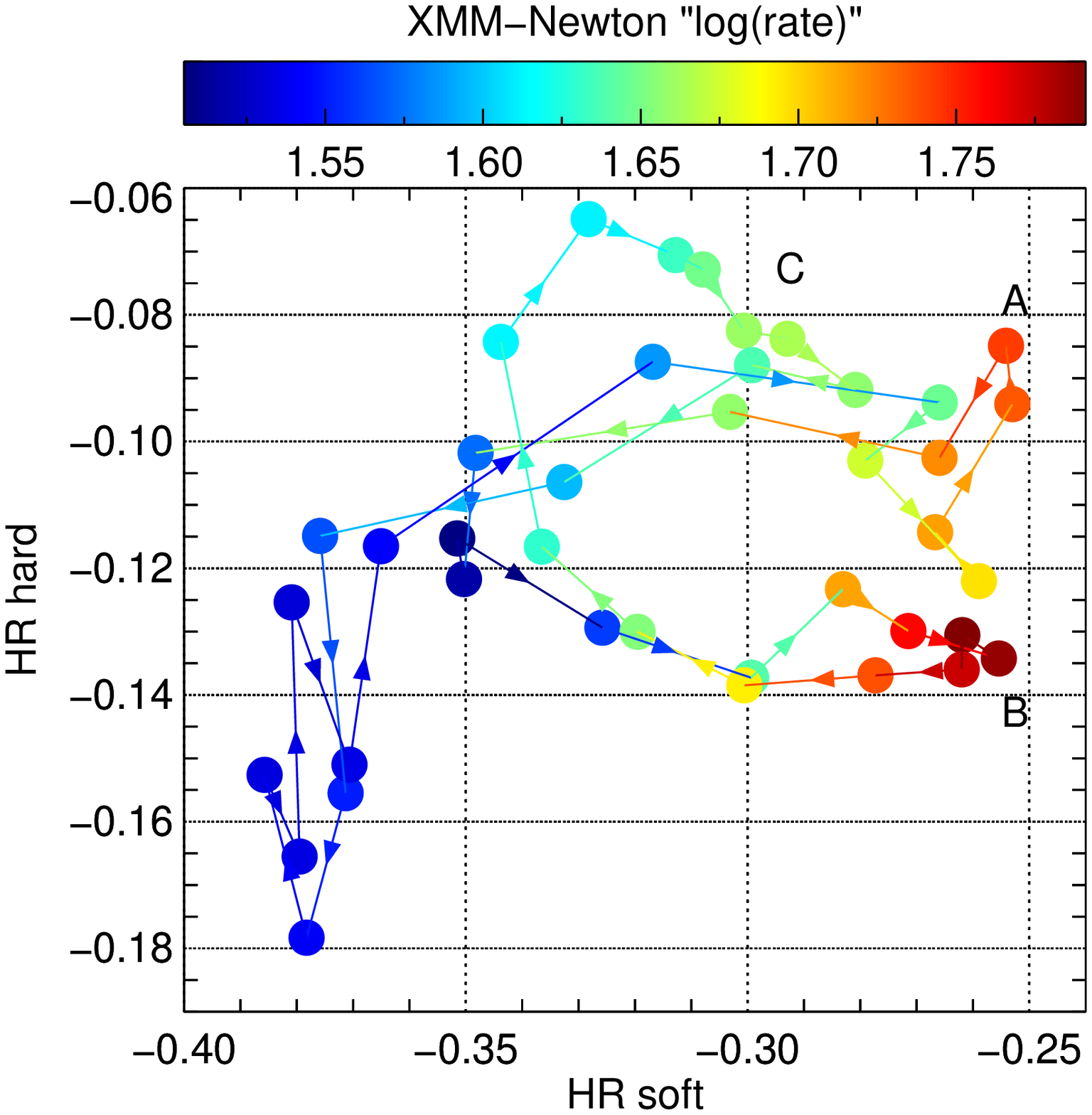}}
  \caption{
    { Left:} Hardness ratio evolution of the average X-ray spectrum of \sxp as estimated for all \swift/XRT observations. Calculations were made using the three bands 0.5-2.0 keV, 2.0-4.5 keV, and 4.5-10.0 keV. HR soft was estimated using the two former energy bands, while HR hard was estimated for the latter two. Colors are representative of the average count rate of each observation, while arrows connect the observations in chronological order.
  { Right:} Hardness ratio evolution of the phase-resolved X-ray spectrum of \sxp based on the \xmm observation. HRs were defined as in the left plot, while 40 phase bins were used. Colors are representative of the average count rate of each phase bin, while arrows connect the observations based on their phase evolution. The labels A, B, and C represent the three peaks in Fig. \ref{fig:pp}.}   
  \label{fig:HR_L}
\end{figure*}

\section{Emission pattern, pulse reprocessing, and the observed PF}
\label{sec:irr}

 To investigate the evolution of the PF with luminosity and to investigate the origin of the soft excess as determined from the phase-resolved analysis, we constructed a toy model for the beamed emission. 
The model assumes that the primary beamed emission of the accretion column is emitted perpendicularly to the magnetic field axis, in a fan-beam pattern \citep{1976MNRAS.175..395B}.
Furthermore, it is assumed that all the emission originates in a point source at a height of ${\sim}2-3\,$km from the surface of the NS. A fraction of the fan emission will also be beamed toward the NS surface, off of which it will be reflected, resulting in a secondary polar beam that is directed perpendicularly to the fan beam \citep[this setup is described in][see also their Fig.~4 ]{2013ApJ...764...49T}.  Partial beaming of the primary fan emission toward the NS surface is expected to occur primarily as a result of scattering  by fast electrons at the edge of accretion column \citep{1976SvA....20..436K,1988SvAL...14..390L,2013ApJ...777..115P}, and to a lesser degree as a result of gravitational  bending of the fan-beam emission. The latter may also result in the emission of the far-away pole entering the observer line of sight. In the model, gravitational bending is accounted for according to the predictions of \cite{2002ApJ...566L..85B}.

The size of the accretion column  (and hence the height of the origin of the fan-beam emission) is a function of the NS surface magnetic field strength and the mass accretion rate \citep[see][ and references within]{2015MNRAS.454.2539M}.
As it changes, it affects the size of the illuminated region on the NS surface and therefore the strength of the reflected polar emission.
If the accretion rate drops below the critical limit, the accretion column becomes optically thin in the direction parallel to the magnetic field axis, and the primary emission is then described by the pencil-beam pattern \cite{1975A&A....42..311B}. In principle, the polar and pencil-beam emission patterns can be phenomenologically described by the same algebraic model. We also modeled the irradiation of any slab or structure (e.g., the inner part of the accretion disk that is located close to the NS) by the combined beams. Last, while arc-like accretion columns and hot-spots have been found in 3D magnetohydrodynamic simulations \citep{roma2004} of accreting pulsars, we did not include them in our treatment as they would merely introduce an anisotropy in the pulse profile and not significantly affect our results.

\subsection{Pulsed fraction evolution} 
The beaming functions of the polar and fan beams are given by $f\sin^m{\phi}$ and $p\cos^k{\phi}$, respectively. Here $\phi$ is the angle between the magnetic field axis and the photon propagation and is thus a function of the angle between the magnetic and rotation axis ($\theta$), the angle of the NS rotation axis, the observing angle ($i$), and the NS spin phase. The exponent values $m$ and $k$ can be as low as 1, but in the general case of the accretion column emission, they can have significantly higher values \citep{2013ApJ...764...49T}.
We constructed various pulse profiles for different combinations of fan- and polar-beam emission patterns with different relative intensities of the fan and polar beams (i.e., $F_{\rm Polar}/F_{\rm Fan}$ ranging from $\sim$0.01 to ${\sim}30$). We note that these values exceed any realistic configuration between the fan beam and the reflected polar beam. More specifically, since the polar beam is a result of reflection of the fan beam, it is only for a very limited range of observing angles that its contribution will (seemingly) exceed that of the fan beam, and thus a value of $F_{\rm Polar}/F_{\rm Fan}\gtrsim2$, is unlikely.
On the other hand, the increasing height of the emitting region, which would result in a smaller fraction of the fan beam being reflected by the NS surface, is limited to $\lesssim10\,km$ \citep{2015MNRAS.454.2539M}, thus limiting the $F_{\rm Polar}/F_{\rm Fan}$ ratio to values greater than 0.1 \citep[see, e.g., Fig 2. of][]{ 2013ApJ...777..115P}. Despite these physical limitations, we have extended our estimates to these exaggerated values in order to better illustrate  the contribution of each component (fan and polar) to the PF. Moreover, a value of $F_{\rm Polar}/F_{\rm Fan}$ 
exceeding 10 qualitatively describes the pencil-beam regime, which is expected at lower accretion rates. We estimated the PF for a range of observer angles and for different combinations of angles between the magnetic and rotation axis. An indicative result for a 45$^o$ angle between the NS magnetic field and rotation axis and for an $r_G/r_{NS}=0.25$ is plotted in Fig~\ref{fig:PF_evol}.  

\subsection{Reprocessed radiation} 
To investigate the origin of the soft excess (i.e., the thermal component in the spectral fits) and to assess its contribution to the pulsed emission,
we studied the pattern of the reprocessed emission from irradiated optically thick material in the vicinity of the magnetosphere. 
More specifically, we considered a cylindrical reprocessing region that extends symmetrically above and below the equatorial plane. The radius of the cylinder is taken to be equal to the magnetospheric radius, where the accretion disk is interrupted.  The latitudinal size of the reprocessing region corresponds to an angular size of 10$^{\rm o}$, as viewed from the center of the accretor \citep[see, e.g.,][and our Fig.~\ref{fig:cher} for an illustration]{2004ApJ...614..881H}. 
The model takes into account irradiation of this region by the combined polar- and fan-beam emission. 
If {\bf d} is the unit vector toward the observer and {\bf n} is the disk surface normal, the total radiative flux that the observer sees would be estimated from their angle {\bf d${\cdot}$n}. We furthermor assume that the observer sees only half of the reprocessing region (and disk, i.e., $\pi$).
Light-travel effects of the scattered/reprocessed radiation \citep[e.g., equation 3 of][]{2016MNRAS.455.4426V} should have a  negligible effect in altering the phase of the reprocessed emission, as the light-crossing time of the disk is less than 1\% (0.05\% $=(1000 km /c)/P_{NS} *100\%$) of the NS spin period.
Thus, any time lag that is expected should be similar in timescale to the reverberation lag observed in LMXB systems \citep[e.g., H1743-322:][]{2016ApJ...826...70D}.

To determine the fraction of the pulsed emission that illuminates the reprocessing region, we integrated the emission over the 10$^{\rm o}$ angular size of the reprocessing region and estimated the PF for an inclination angle between the rotational and the magnetic filed axes ($\theta$) that ranges between 0$^{\rm o}$ and 90$^{\rm o}$ (Fig.~\ref{fig:irrad}, dotted line). 
We also estimate the PF originating from the reprocessing region assuming a ratio of  $F_{\rm Polar}/F_{\rm Fan}=$0.25 (Fig.~\ref{fig:irrad}, solid line). To further probe the contribution of the two beam components to the reprocessed emission, we estimated the PF for two extreme cases in which all the primary emission is emitted in the polar  beam (Fig.~\ref{fig:irrad}, green dashed line) or only the fan beam (Fig.~\ref{fig:irrad}, blue dashed line).  As we discussed, this scenario is unrealistic, since the polar beam is the result of reflection of the fan beam. However, this setup would adequately describe the pencil-beam emission, expected at lower accretion rates \cite[e.g.,][]{1975A&A....42..311B}.

\begin{figure}
   \resizebox{\hsize}{!}{\includegraphics[angle=0,clip=, bb=13 10 538 425 ]{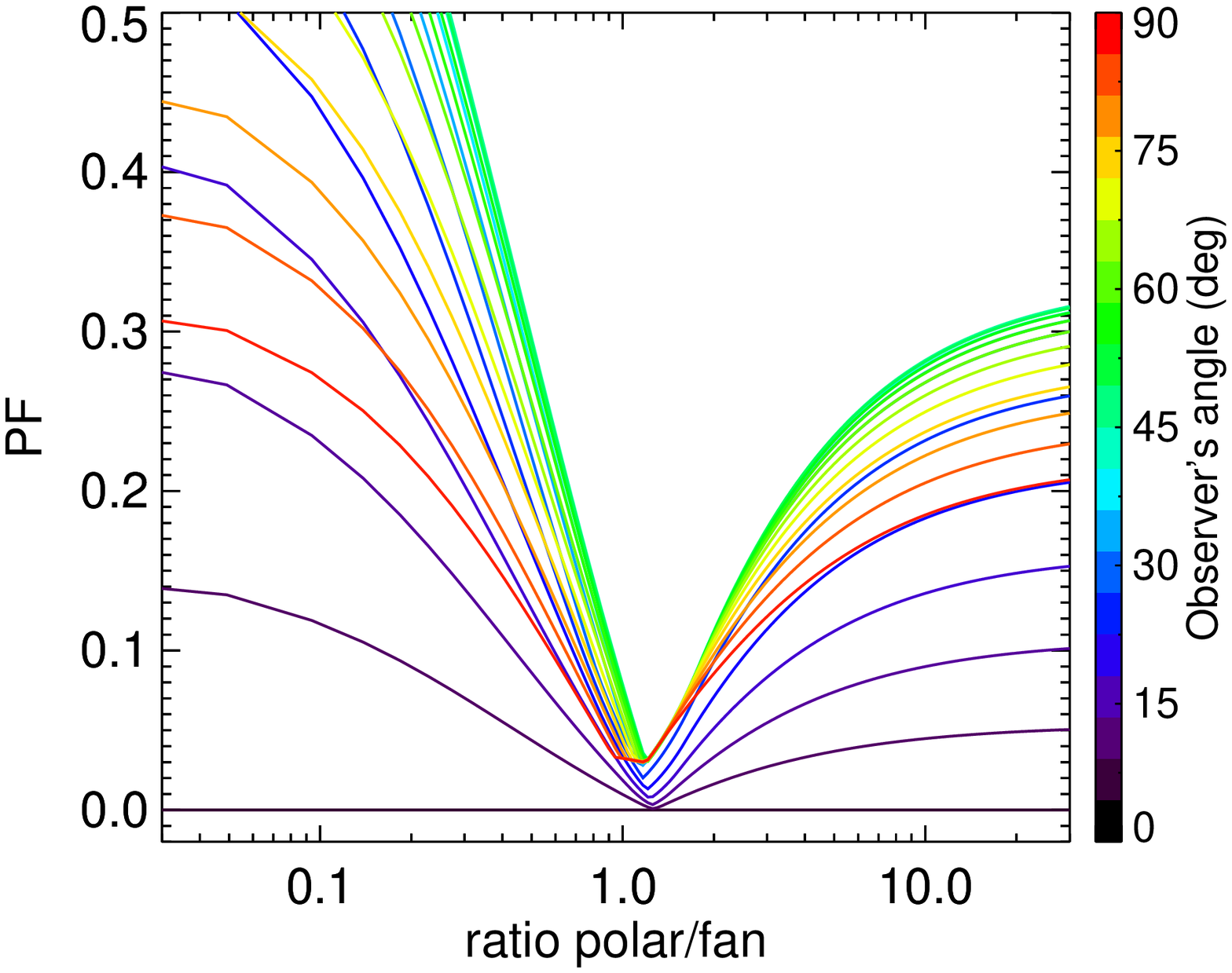}}
   \caption{PF versus intensity between fan and pencil beam.}
   \label{fig:PF_evol}
\end{figure}

\begin{figure}
   \resizebox{\hsize}{!}{\includegraphics[angle=0,clip=, bb= 0 0 512 400]{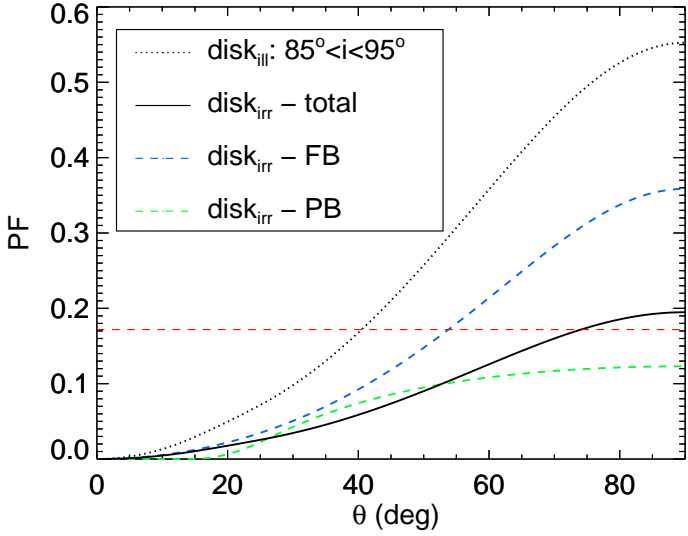}}
   \caption{PF of the reprocessed emission of a fan-beam pattern for various inclination angles between the magnetic and rotation axis. 
   The dotted line represents the PF of the integrated emission over a cylindrical surface of angular size of 10$^{\rm o}$ and located at the equatorial plane of the NS. The solid line is the calculated reprocessed PF for a combination of polar beam
and fan beam ($F_{\rm Polar}/F_{\rm Fan}=$0.25). The dashed blue and green lines show the PF of reprocessed emission that would originate from a pure fan-beam and polar-beam illuminating pattern, respectively. The red horizontal line represents the limit on the PF given the variability of the observed flux of the thermal component, presented in Figure~\ref{fig:phace_spec_res}. }
   \label{fig:irrad}
\end{figure}  

 \begin{figure}
   \resizebox{\hsize}{!}{\includegraphics[angle=0,clip=,bb=11 11 500 400]{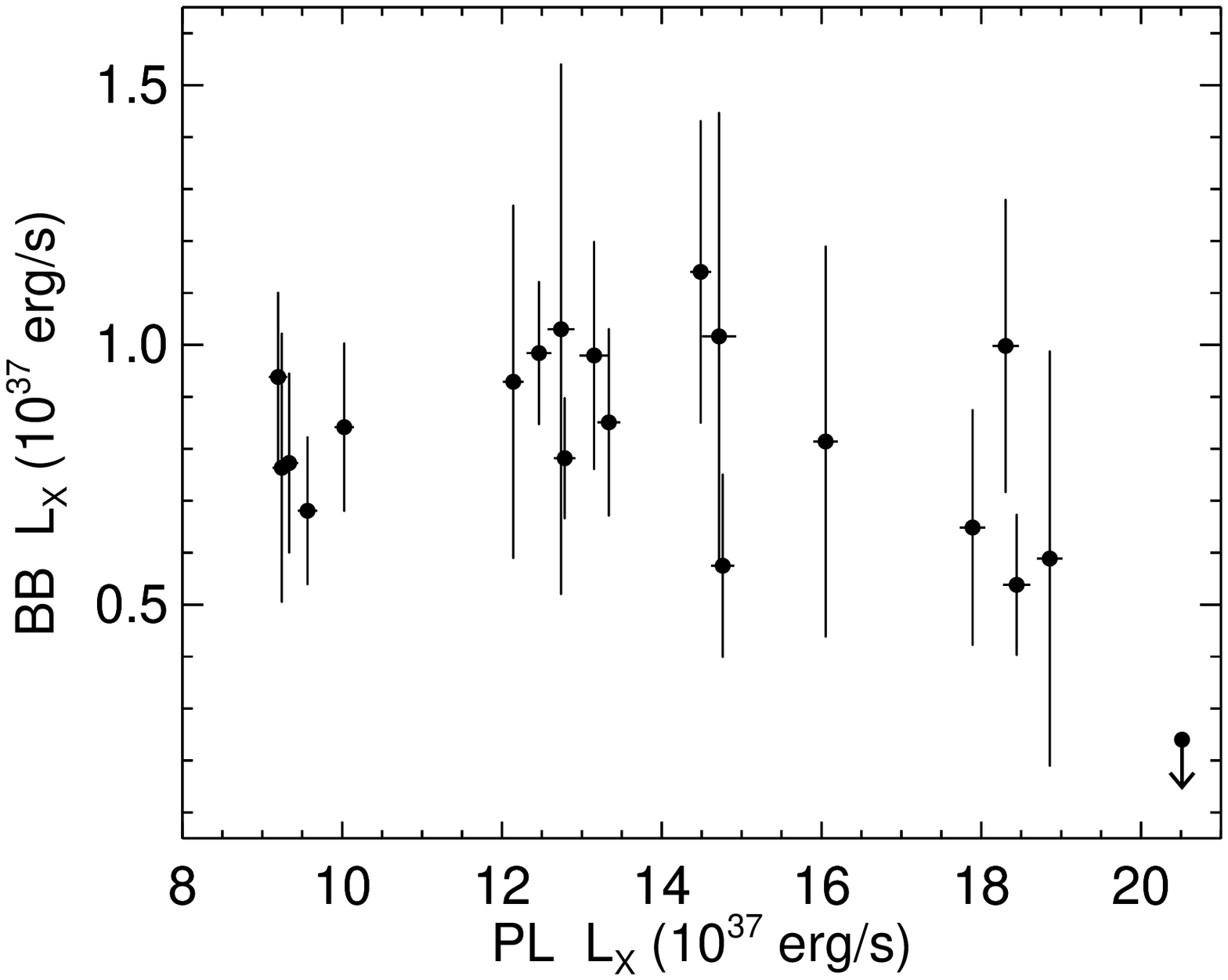}}
   \caption{Absorption-corrected PL X-ray luminosity vs bolometric disk-BB luminosity.
   No clear correlation between the two quantities was established.
   }
   \label{fig:L_BB_PL}
\end{figure}

\section{Discussion}
 \label{discussion}

 The high quality of the \xmm spectrum of SMC-X3 and the very large number of total registered counts have allowed us to perform a detailed time-resolved spectroscopy where we robustly constrained the different 
 emission components and monitored their evolution with pulse
phase. Furthermore, the high time-resolution of the \xmm detectors, complimented with the numerous Swift/XRT observations (conducted throughout the outburst), have allowed for a detailed study of the long- and short-term temporal behavior of the source. Below, we present our findings and their implications with regard to the underlying physical mechanisms that are responsible for the observed 
 characteristics.

\subsection{Emission pattern of the accretion column and its evolution}

The 0.01-12\,keV luminosity of SMC~X-3 during the \xmm observation is estimated at 1.45$\pm0.01\times10^{38}$\,erg/s for a distance of ${\sim}$62\,kpc. The corresponding mass accretion rate, assuming an efficiency of $\xi=0.21$ \citep[e.g.,][]{2000AstL...26..699S}, is ${\dot M}{\sim}1.2\times10^{-8}\,{M_ \odot }/{\rm yr}$, which places the source well within the fan-beam regime \citep{1976MNRAS.175..395B,1981ApJ...251..288N}. Furthermore, the shape of the pulse profile (Fig.~\ref{fig:pp}) indicates a more complex emission pattern, comprised of the primary fan beam and a secondary reflected polar beam, as was proposed in \cite{2013ApJ...764...49T} and briefly presented in Section~\ref{sec:irr}. The energy-resolved pulse profiles and the phase-resolved hardness ratios presented in Fig.~\ref{fig:HR} strongly indicate that the pulsed emission is dominated by hard-energy photons, while  the smooth single-color appearance of the 0.2-0.6\,keV energy range in the lower bin heat map presented in Fig.~\ref{fig:phase_heat} indicates a soft spectral component that does not pulsate.

Using our toy-model for the emission of the accretion column, we were able to explore the evolution of the PF with the $F_{\rm Polar}/F_{\rm Fan}$ ratio, assuming a  broad range of observer viewing angles. Interestingly, the resulting PF vs $F_{\rm Polar}/F_{\rm Fan}$ ratio (Fig.~\ref{fig:PF_evol}) qualitatively resembles the evolution of the PF, with increasing source count-rate, as presented in Fig.~\ref{fig:cr_pf}, using Swift/XRT data. Following the paradigm of \cite{1975A&A....42..311B} and \cite{1976MNRAS.175..395B}, we expect the emission pattern of the accretion column to shift from a pencil- to a fan-beam pattern as the accretion rate increases. Below $\sim$1\,cts/sec, the emission is dominated by the pencil beam (which is qualitatively similar to the polar beam). This regime is defined by $F_{\rm Polar}/F_{\rm Fan}>2$.  As the luminosity increases, the radiation of the accretion column starts to be emitted in a fan beam, which is accompanied by the secondary polar beam, corresponding to $0.7{\lesssim}F_{\rm Polar}/F_{\rm Fan}{\lesssim}2$ (this only refers to the X-axis of Fig.~\ref{fig:cr_pf} and does not imply chronological order, which is indicated with arrows in Fig.~\ref{fig:HR_L}). This is the era during which the \xmm observation was conducted. As the accretion rate continued to increase, the accretion column grew larger, the height of the emitting region increased, and the emission started to become dominated by the fan beam as the reflected component decreases, yielding a lower value of the $F_{\rm Polar}/F_{\rm Fan}$ ratio.
Nevertheless, it is evident from Fig.~\ref{fig:cr_pf}  that the PF does not readily increase as the fan beam becomes more predominant (i.e., above ${\sim}7$\,cts/s). This indicates that the height of the accretion column is limited to moderate values \citep[e.g., as argued by][]{2013ApJ...777..115P,2015MNRAS.454.2539M}, and therefore a considerable contribution from the polar beam is always present.

The presence of the polar beam is also indicated by the shape of the pulse profile and its evolution during different energy bands and HR values (Figures~\ref{fig:pp} and \ref{fig:HR}). The pulse profile has three distinct peaks that we label A, B, and C from left to right in Fig.~\ref{fig:pp}. The energy and HR-resolved pulse profiles presented in Fig.~\ref{fig:HR} reveal that the pulse peaks are dominated by harder emission. A closer inspection indicates that peak B is softer than the other two peaks and is still present (although much weaker) in the lower energy bands (Fig.~\ref{fig:HR}, lower left bin). This finding is further supported by our study of the HR evolution of the sources throughout the \xmm observation, presented in Fig.~\ref{fig:HR_L} (right panel).  This representation clearly shows that the source is harder during high-flux intervals (the peaks of the pulse profile), but also that peak B is distinctly softer than peaks A and C. We argue that these findings indicate a different origin of the three peaks. More specifically, we surmise that the softer and more prominent peak B originates mostly from the fan beam, while the two harder peaks (A and C) are dominated by the polar-beam emission, which is expected to be harder than the fan beam. This is because only the harder incident photons (of the fan beam) will be reflected (i.e., backscattered) off the NS atmosphere, while the softer photons will be absorbed \citep[e.g.,][]{2013ApJ...764...49T,2015MNRAS.452.1601P}.

 \begin{figure}
   \resizebox{\hsize}{!}{\includegraphics[angle=0]{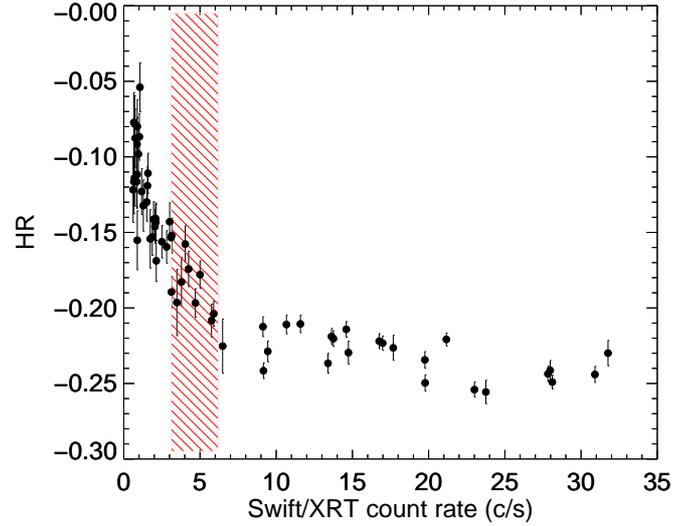}}
   \caption{Hardness ratio of the 1-3\,keV to 3-10\,keV bands vs count rate for the \swift observations of SMX~X-3 between MJDs 57611 and 57754. The red-striped area denotes the luminosity correspoding to 1-2 Eddington rates (0.3-10 keV).
   }
   \label{fig:HR_swift}
\end{figure}  

The analysis of the \swift data taken throughout the outburst allowed us to also probe the long-term spectral evolution of the source. In the left panel of Fig.~\ref{fig:HR_L} we present the HR ratio evolution of the source throughout the 2016 outburst. Unlike the short-term evolution of the source, during which the emission is harder during the high-flux peaks of the pulses (see the data from the \xmm observation presented in the right panel of Fig.~\ref{fig:HR_L}), the emission of SMC~X-3 becomes softer in the long term as the luminosity of the source increases (color-coded count rate in Fig.~\ref{fig:HR_L}). This behavior is also evident in Fig.~\ref{fig:HR_swift}, where the HR ratio of the 1-3\,keV to the 3-10\,keV band is plotted versus the source count-rate. The source becomes clearly softer as the accretion rate (which corresponds to the observed flux) increases.  Long-term observations of numerous XRPs suggests that they can be divided into two groups, based on their long-term spectral variability. Sources in group 1 (e.g., 4U~0115+63) exhibit a positive correlation between source hardness and its flux, while sources in group 2 (e.g., Her X-1 or this source) a negative \citep[e.g.,][]{2007AstL...33..368T,2010MNRAS.401.1628T,2011A&A...532A..99V,2011A&A...532A.126K}. 

It can be argued that the apparent two populations are just the result of observing the XRPs in two different regimes of mass accretion rates. At low accretion rates, deceleration of the in-falling particles in the accretion column occurs primarily via Coulomb interactions, while in the high-rate regime, the mass is decelerated through pressure from the strong radiation field \citep[e.g.,][]{1976MNRAS.175..395B,1981A&A....93..255W,2015MNRAS.454.2539M}.  As   the mass accretion rate increases, so does the height of the accretion column, and therefore the reflected fraction (i.e., the polar beam) starts to decrease. As the polar emission is harder than the fan beam, the total registered spectrum stops hardening and even becomes moderately softer \citep[][]{2015MNRAS.452.1601P}. Observationally, the positive correlation is expected at luminosities below a critical value and the negative above it. This critical value is theoretically predicted at ${\sim}1-7\times10^{37}$\,erg/s \citep[e.g.,][]{2007A&A...465L..25S,2011A&A...532A.126K,2012A&A...544A.123B,2015MNRAS.452.1601P}, which is in agreement with the earlier prediction of \cite{1976MNRAS.175..395B}. Observations of XRPs at different luminosities indeed show that a single source can exhibit the behavior of both groups when monitored above or below this critical value \citep[e.g.,][]{2015MNRAS.452.1601P}.

The high-luminosity regime, which is the only regime probed by the Swift data points in Fig.~\ref{fig:HR_swift},  is essentially the diagonal branch of the hardness-intensity diagrams presented by \cite{2013A&A...551A...1R}.
The source behavior, presented in Fig.~\ref{fig:HR_swift}, is consistent with the theoretical predictions for sources accreting above the critical luminosity of ${\sim}10^{37}$\,erg/s.  However, while the previous observations by \cite{2011A&A...532A.126K} and \cite{2015MNRAS.452.1601P} showed the pause in the hardening of the source with increasing luminosity, and (in some sources)  the slight decrease in hardness, this drop is clearly demonstrated
here thanks to the numerous high-quality observations provided by the { Swift} telescope. Furthermore, our analysis indicates the presence of perhaps a third branch that appears when the source luminosity exceeds ${\sim}4{\times}L_{\rm Edd}$ , in which the source luminosity increases but the emission does not become softer. If this additional branch is indeed real, it may have eluded detection by \citeauthor{2013A&A...551A...1R}  simply because the sources studied there never reached such high luminosities. It is plausible that the stabilization of the source hardness is a manifestation of the predicted physical limitations imposed on the maximum height of the accretion column \citep[e.g.,][]{2013ApJ...777..115P,2015MNRAS.454.2539M}. }

\subsection{Origin of the soft excess and the optically thin emission}

The phase-averaged spectral continuum of SMC~X-3 during the \xmm observation is described by a combination of a emission in the
shape of a hard power-law and a softer thermal component with a temperature of ${\sim}0.24$\,keV. The nonthermal component is consistent with emission from the accretion column \cite[e.g.,][]{2007ApJ...654..435B}, but the origin of the soft thermal emission is less clear. Assuming that the accretion disk is truncated approximately at the magnetosphere, and for an estimated magnetic field of ${\sim}2-3\times10^{12}$\,G \citep{2014MNRAS.437.3863K,2017arXiv170200966T}, the maximum effective temperature of the accretion disk would lie in the far-UV range (20-30\,eV) for ${\dot M}{\sim}1.2\times10^{-8}\,{M_ \odot }/{\rm yr}$, as inferred from the observed luminosity and for $R_{\rm in}=0.5-1\,R_{\rm mag}$. 
Although we can exclude a typical thin disk as the origin of the emission, an inflated disk that is irradiated from the NS might be the origin of the thermal emission, because such a disk can reach higher temperatures \citep[see, e.g.,][and the discussion below]{2017arXiv170307005C}.

The emission is also unlikely to originate in hot plasma on the surface of the NS, as the size of a blackbody-emitting sphere that successfully fits the data is an order of magnitude larger (${\sim}170$\,km) than the ${\sim12}$\,km radius of a 1.4\,${M_ \odot }$ NS. The observed thermal emission could be the result of reprocessing of the primary emission by optically thick material that lies at a distance from the NS and is large enough to partially cover the primary emitting region (i.e., the accretion column). If a significant fraction of the primary emission is reprocessed, then this optically thick region will emit thermal radiation at the observed temperatures. A detailed description of this configuration is presented in \cite{2000PASJ...52..223E} and \cite{2004ApJ...614..881H}. This optically thick material is composed of accreted material trapped inside the magnetosphere and also the inside edge of the accretion disk, which at these accretion rates will start to become radiation dominated and geometrically thick \citep[e.g.,][]{1973A&A....24..337S,2007ARep...51..549S,2017arXiv170307005C}. The reprocessing region is expected to have a latitudinal temperature gradient, and its emission can be modeled as a multi-temperature blackbody \citep[e.g.,][]{2017MNRAS.467.1202M}, which we modeled using the {\small XSPEC}  model {\texttt{diskbb}}. Nevertheless, we stress that the soft emission most likely does not originate in an accretion disk heated through viscous dissipation,  but is rather the result of illumination of an optically thick region at the disk-magnetosphere boundary. Following \citeauthor{2004ApJ...614..881H} and assuming that the reprocessing region subtends a solid angle $\Omega$ (as viewed from the source of the primary emission) and has a luminosity given by $L_{\rm soft}$ = ($\Omega$/4$\pi$) $L_{\rm X}$, where  $L_{\rm X}$ is the observed luminosity and further assuming that the reprocessed emission is emitted  isotropically and follows a thermal distribution (i.e. $L_{\rm soft} = \Omega R^2 \sigma T^4_{\rm soft}$), we can estimate the distance between the primary emitting region and the reprocessing region from the relation $R^2 = L_{\rm X}$/($4\pi \sigma T^4_{\rm soft})$ \citep{2004ApJ...614..881H}. 

For our best-fit parameters and for a spectral hardening factor of $\sim1.5-1.7$ \citep{1995ApJ...445..780S}, the relation yields a value of $1.5{\pm}0.3{\times}10^{8}$\,cm, which is in very good agreement with the estimated magnetospheric radius of $R_{\rm m}=1.6{\pm}0.4{\times}10^{8}$\,cm that corresponds to the expected (i.e., ${\sim}2-3{\times}10^{12}$\,G) magnetic field strength of SMC~X-3 and for 
$R_{\rm m} {\sim}2 \times 10^7{\alpha}{\dot{M}_{15}}^{-2/7}{B_{9}}^{4/7}{M_{1.4}}^{-1/7}R_6^{12/7}$ cm \citep{1977ApJ...217..578G,1979ApJ...232..259G,1979ApJ...234..296G,1992xbfb.work..487G}, where ${\alpha}$  is a constant that depends on the geometry of the accretion flow, with ${\alpha}$=0.5 the commonly used value for disk accretion, ${\dot{M}_{15}}$ is the mass accretion rate in units of $10^{15}$\,g/s (estimated for the observed luminosity of ${\sim}$1.5${\times}10^{38}$\,erg/s and assuming an efficiency of 0.2), $B_{9}$ is the NS magnetic field in units of $10^{9}$ G, $M_{1.4}$ is the NS mass in units of 1.4 times the solar mass, and $R_6$ is the NS radius in units of $10^6$ cm.

Three prominent broad emission lines are featured on top of the broadband continuum. Centered at ${\sim}$0.51\,keV ${\sim}$0.97\,keV and ${\sim}$6.64\,keV, the lines are consistent with emission from hot ionized plasma. The 6.6\,keV line is  most likely due to K-shell fluorescence from highly ionized iron, while the 0.5\,keV line is most likely the K${\alpha}$ line of the  \ion{N}{VII} - Ly$\alpha$ (500) ion. The 1\,keV line is most likely the result of combined Ne K${\alpha}$ and Fe L-shell emission lines. Both the 1\,keV and 6.6\,keV lines appear to be broadened,  although as we discuss below, the broadening is most likely artificial and is the result of blending of unresolved emission lines from relevant atoms at different ionization states. Both the 1\,kev and the 6.6\,keV lines are often detected in the spectra of XRBs and appear to be particularly pronounced in the spectra of X-ray pulsars (e.g., \citealt{2003A&A...407.1079B}; \citealt{2006A&A...445..179D}; \citealt{2010A&A...522A..96N}; \citealt{2014MNRAS.437..316K}). While in many XRBs the emission lines (and particularly the iron K${\alpha}$ line) are attributed to reflection of the primary emission from the optically thick accretion disk \cite[e.g.,][and references therein]{2010LNP...794...17G}, in the case of X-ray pulsars (and in this source), the emission line origin is more consistent with optically thin emission from rarefied hot plasma that is trapped in the Alfv{\'e}n shell of the highly magnetized NS \cite[e.g.,][]{1978ApJ...223..268B}. The (apparent) line broadening is consistent with microscopic processes, i.e., Compton scattering for a scattering optical depth of 0.5-1 \citep[e.g., equation 30 in][]{1978ApJ...223..268B}, rather than macroscopic motions, such as the rotation of the accretion disk or the surface of the donor star. The measured width of the two lines corresponds to line-of-sight velocities of ${\sim}1.6\times10^{4}$\,km/sec. For the disk inclination of $\lesssim45^o$, expected for SMC X-3 \citep{2017arXiv170102336T}, this corresponds to a 3D velocity of $\gtrsim2.3\times10^{4}$\,km/sec. Assuming Keplerian rotation, these velocities would correspond to a distance of $\lesssim350$\,km from the NS, which is almost an order of magnitude smaller than the estimated magnetospheric radius. Furthermore, the phase-resolved pn spectra and the RGS spectra (discussed below) indicate that the line broadening may be an artifact, resulting from the blending of unresolved thin emission lines at different centroid energies.

\subsection{Short-term evolution of the spectral continuum and emission lines. }

The timing analysis of the source emission has revealed that the pulsed emission is dominated by hard photons (Fig.~\ref{fig:phase_heat}), and the phase-folded light-curves additionally revealed a non-pulsating component of the total emission at energies below $\sim1$\,keV (see Fig.~\ref{fig:HR} and Fig.~\ref{fig:phase_heat}, bottom). 
The soft persistent component of the source light-curve appears to coincide with the thermal spectral component. To investigate this further, we divided the pn spectra into 20 equally timed phase intervals, which were modeled  independently. We find that while the contribution from the soft thermal component is more or less stable (Fig.~\ref{fig:L_BB_PL}), the contribution of the power-law component strongly correlates with the pulse profile (Fig.~\ref{fig:phace_spec_res}). We note, however, that the use of a simple power law in our spectral fittings fails to model the low-energy roll-over expected from a photon distribution, which is the result of multiple Compton upscattering of soft, thermal seed photons. Nevertheless, adding a parameter for the low-energy turn-over would only add to the uncertainties of the soft component parameters, but would not qualitatively affect these findings. The variation of the power-law component is also illustrated in Fig.~\ref{fig:spec_res}, in which the phase-resolved spectra were unfolded from the phase-averaged model. This behavior further supports the presence of a large, optically thick reprocessing region located at a considerable distance (i.e., the magnetospheric boundary). As demonstrated in Fig.~\ref{fig:irrad}, the reprocessed emission contributes a very small fraction of the pulsed emission.

This finding contradicts the conclusions of \cite{2009A&A...498..217M} regarding another known Be-XRB, XMMU J054134.7-682550. \citeauthor{2009A&A...498..217M} also noted thermal emission in the spectrum of the source (they simultaneously analyzed \xmm and RXTE data), which they also attributed to reprocessing of the primary emission by optically thick material. However, contrary to this work, \citeauthor{2009A&A...498..217M} concluded that the reprocessed emission also pulsates. They furthermore argued that this is the result of highly beamed emission from the pulsar that is emitted toward the inner disk border. The authors reached this conclusion by noting the (tentative) presence of a sinusoidal-like shape of the pulse profile in the $\lesssim$1\,keV range. They then concluded that this is mostly due to the contribution of the reprocessed emission. 
However, the apparent pulsations of the soft component may be due to the fact that \citeauthor{2009A&A...498..217M} did not distinguish between the 0.5-1\,keV and  ${<}0.5$\,keV energy range, as the total number of counts in their observation did not allow it. It is highly likely that the authors were probing the lower energy part of the accretion column emission and not the reprocessed thermal emission. In our Fig.~\ref{fig:HR} (left), peak  B (and to a lesser extent, peak A) is still detected in the 0.5-1\,keV range and it is only below ${\sim}0.5$\,keV (where the thermal component dominates the emission) that the pulses disappear. More importantly, the high quality of our observation allowed us to perform the detailed pulse-resolved analysis that demonstrated the invariance of the thermal component.

Furthermore, the fan-beam emission is not expected to be extremely narrow and while optically thick material in the disk/magnetosphere boundary is expected to become illuminated by the primary emission (as both this work and \citealt{2009A&A...498..217M} propose), there is no reason to expect that the beam  will be preferentially (and entirely) aimed toward the disk inner edge (especially when one accounts for the gravitational bending). Of course, XMMU J054134.7-682550 could be a special case in which such an arrangement took place. However, we conclude that the higher quality of the data available for \sxp most likely allowed for a more detailed analysis that revealed the non-pulsating nature of the thermal emission. This finding highlights the importance of long-exposure high-resolution observations of X-ray pulsars.

The emission lines noted in the phase-averaged spectra are also detected in (most of) the phase-resolved spectra.  Their apparent strength and centroid energy variation between different phases (when the lines are detected, see Table~\ref{tab:lines}) lies within the 90\% confidence range. However, the lines completely disappear during some phase intervals.  While there is no discerning pattern between the presence (or absence) of the lines and the pulse phase, this variability is striking. Furthermore, it is strongly indicated (from the RGS data) that the seemingly broadened lines of the phase-averaged pn spectrum are the result of blending of narrow emission features at different centroid energies. The presence and variability of multiple emission lines indicates optically thin material that is located close to the central source and has a complex shape. As the different parts of this structure are illuminated by the central source, their ionization state varies with phase and position, resulting in emission lines with moderately different energies and different strengths. Such a complex structure is also supported by the predictions of \citep{roma2004} for accretion of hot plasma onto highly magnetized NSs.

 Based on our line identification in the RGS data, no significant blueshift can be established for the emission lines listed in Table~\ref{tab:rgs}.
We note that similar lines have been reported in the X-ray spectra of other BeXRBs during outbursts \citep[e.g., SMC\,X-2, SXP\,2.16][]{2016MNRAS.458L..74L,vas2017}. For SMC\,X-2, which was observed by \xmm during a luminosity of $\sim$1.4\ergs{38} (0.3-12.0 keV band), it has been proposed that these lines could be associated with the circum-source photoionized material in the reprocessing region at the inner disk \citep{2016MNRAS.458L..74L}.  However, if the apparent emission line variability noted in out analysis is real, this would indicate a complex structure of optically thin plasma that lies closer to the source of the primary emission and is illuminated by it at different time intervals. Photoionized material at the boundary of the magnetosphere would not exhibit such rapid variability.

\subsection{SMC~X-3 in the context of ULX pulsars }

It is interesting to note that all the emission lines resolved in \sxp have also been identified in ultraluminous X-ray sources \citep[NGC 1313 X-1 and NGC 5408 X-1,][]{2016Natur.533...64P,2017AN....338..234P} and ultraluminous soft X-ray sources \citep[NGC\,55\,ULX, ][]{2017MNRAS.468.2865P}. Moreover, there are striking similarities between the 1\,keV emission feature of \sxp and the same feature as described for NGC\,55\,ULX within the above studies; ``emission peak at 1 keV and absorption-like features on both sides'' \citep{2017MNRAS.468.2865P}. However, we note that for \sxp and other BeXRB systems, to our best knowledge, no significant blueshift of these lines has been measured. This might mean that these lines are produced by the same mechanism both in BeXRB systems (observed at luminosities around the Eddington limit) and in ULXs. However, there is no ultrafast wind/outflow in BeXRBs, as in the case of ULX systems.

\begin{figure}
   \resizebox{\hsize}{!}{\includegraphics[angle=0,clip=0]{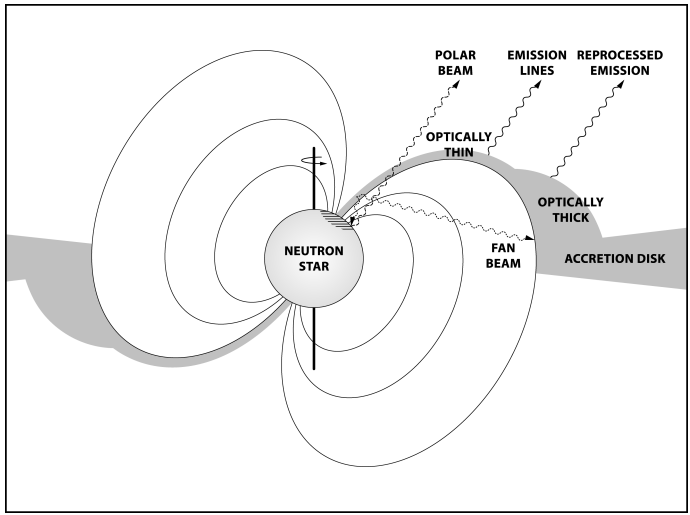}}
   \caption{Schematic representation of the proposed structural configuration of X-ray pulsars in the moderately high accretion regime. The (possibly inflated) accretion disk is truncated at approximately the magnetosphere, and a combination of optically thick and thin plasma is trapped at the magnetopsheric boundary. The optically thick material (and the inner disk edge) partially cover the primary emission source and reprocess its emission. At higher latitudes, the material remains optically thin. This is the origin of the emission lines. The source of the primary emission is located at a some height above the NS pole, inside the accretion column.  We stress that the drawing is aimed to better illustrate the discussed configuration and does not attempt to realistically reproduce the geometry of an accreting highly magnetized NS. More specifically, for viewing clarity, the emission pattern and direction of the fan and polar beams has been oversimplified and
   the inner disk radius (and size of the magnetosphere) has been greatly depreciated (e.g., in a more realistic example, both components may illuminate the reprocessing region). The regions where the optically thin and thick plasma are expected to lie have also been presented in a very simplified minimalistic arrangement.
   }
   \label{fig:cher}
\end{figure}

The findings of this work regarding the continuum and line emission and their temporal variations construct a picture of X-ray pulsars in which the accretion disk is truncated at a large distance from the NS (i.e., the magnetosphere), after which accretion is governed by the magnetic field. As material is trapped by the magnetic field, an optically thick structure is formed inside the pulsar magnetosphere, which (for a range of viewing angles) covers the primary source, partially reprocessing its radiation. The optically thick region is expected to have an angular size on the order of${\text{}}$ a few tens of degrees and to be centered at the latitudinal position of the accretion disk. At higher latitudes and closer to the accretion column, the trapped plasma remains optically thin, producing the observed emission lines. This configuration is schematically represented in Fig.~\ref{fig:cher} and is in agreement with similar schemes proposed by  \cite{2000PASJ...52..223E} and \cite{2004ApJ...614..881H}.  At high accretion rates, the size of the reprocessing region becomes large enough to reprocess a measurable fraction of the primary emission. This is due to the increase in the accreting material trapped by the pulsar magnetosphere and the increase in the thickness of the accretion disk, which in turn is  due to the increase in radiation pressure \citep[e.g.,][]{1973A&A....24..337S, 2007MNRAS.377.1187P,2011MNRAS.413.1623D,2017arXiv170307005C}. 

This optically thick reprocessing region has been noted in numerous X-ray pulsars (e.g., Cen~X-3: \citealt{2000ApJ...530..429B}; Her~X-1: \citealt{2002MNRAS.337.1185R}; LXP~8.04: Vasilopoulos et al. in prep; SMC~X-1: \citealt{2005ApJ...633.1064H}; SMC X-2: \citealt{2016MNRAS.458L..74L}). Its presence in sub- or moderately super-Eddington accreting sources becomes especially pertinent in light of recent publications that postulated that at even higher accretion rates (corresponding to luminosities exceeding $10^{39}$\,erg/s, i.e., in the ULX regime), the entire magnetosphere becomes dominated by optically thick material, which reprocesses all the primary photons of the accretion column \citep[][]{2017MNRAS.467.1202M}, essentially washing out the pulsation information and the spectral characteristics of the accretion column emission. The prediction of this model has also been used to argue that a considerable fraction of ULXs may in fact be accreting, highly magnetized NSs and not black holes \citep{kolio2017}. In light of these considerations, \sxp is of particular significance, since it stands right at the threshold between sub-Eddington X-ray pulsars and ULXs, and its spectro-temporal characteristics support the notion of a reprocessing region that is already of considerable size.

\section{Conclusion}
\label{conclusion}   

We have analyzed the high-quality \xmm observation of \sxp during its recent outburst. The \xmm data where complemented with \swift/XRT observations, which were used to study the long-term behavior of the 
source. By carrying out a detailed temporal and spectral analysis (including phase-resolved spectroscopy) of the source emission, we found that its behavior and temporal and spectral characteristics fit the theoretical expectations and the previously noted observational traits of accreting highly magnetized NSs at high accretion rates. More specifically, we found indications of a complex emission pattern of the 
primary pulsed radiation, which most likely involves a combination of a fan-beam emission component directed perpendicularly to the magnetic field axis and a secondary polar-beam component reflected off the
NS surface and directed perpendicularly to the primary fan beam (as discussed in, e.g., \citealt{1976MNRAS.175..395B}; \citealt{1981A&A....93..255W} and more recently by \citealt{2013ApJ...764...49T}). 

The spectroscopic analysis of the source further reveals optically thick material located at approximately the boundary of the magnetosphere. The reprocessing region has an angular size (as viewed from the NS) that is large enough to reprocess a considerable fraction of the primary beamed emission, which is remitted in the form of a soft thermal-like component that contributes very little to the pulsed emission.  These findings are in agreement with previous works on X-ray pulsars \citep[e.g.,][]{2000PASJ...52..223E,2004ApJ...614..881H}, but also with the theoretical predictions for highly super-Eddington accretion onto highly magnetized NSs \citep[e.g.,][]{2017MNRAS.468L..59K,2017MNRAS.467.1202M,kolio2017}, where it has been argued that this reprocessing region grows to the point that it may reprocess the entire pulsar emission.

\section{Acknowledgements}

The authors thank Maria Petropoulou  for valuable advice and stimulating discussion and Olivier Godet for contributing to the RGS analysis. The authors also extend their warm gratitude to Cheryl Woynarski and woyadesign.com for designing the source schematic. Finally we extend our gratitude to the anonymous referee, whose keen observations significantly improved our manuscript.

\bibliography{general}

\end{document}